\documentclass[showpacs,twocolumn,superscriptaddress,nofootinbib]{revtex4-1} 

\usepackage[utf8]{inputenc}
\usepackage{hyperref}
\hypersetup{colorlinks=true,linkcolor=blue,citecolor=cyan}
\usepackage{graphicx}
\usepackage{xcolor, soul}
\sethlcolor{green}
\usepackage{dcolumn}
\usepackage{bm}
\usepackage{color}
\usepackage{enumitem}
\usepackage{mathrsfs}
\usepackage{amsmath}
\usepackage{amssymb}
\usepackage{cleveref}

\usepackage{soul,xcolor}
\setstcolor{red}

\begin{document}

\title{Kerr-Newman-modified-gravity black hole's impact on the magnetic reconnection}

\author{Sanjar Shaymatov}
\email{sanjar@astrin.uz}
\affiliation{Institute for Theoretical Physics and Cosmology, Zhejiang University of Technology, Hangzhou 310023, China}
\affiliation{Institute of Fundamental and Applied Research, National Research University TIIAME, Kori Niyoziy 39, Tashkent 100000, Uzbekistan}
\affiliation{Western Caspian University, Baku AZ1001, Azerbaijan}

\author{Mirzabek Alloqulov%,\orcidlink{0000-0001-5337-7117}
}
\email{malloqulov@gmail.com}
\affiliation{Institute of Fundamental and Applied Research, National Research University TIIAME, Kori Niyoziy 39, Tashkent 100000, Uzbekistan}
\affiliation{University of Tashkent for Applied Sciences, Gavhar Street 1, Tashkent 100149, Uzbekistan}

\author{Bobomurat Ahmedov}
\email{ahmedov@astrin.uz}
 \affiliation{Institute of Fundamental and Applied Research, National Research University TIIAME, Kori Niyoziy 39, Tashkent 100000, Uzbekistan}

\author{Anzhong Wang}
\email{anzhong\_wang@baylor.edu}
\affiliation{GCAP-CASPER, Physics Department, Baylor University, Waco, Texas 76798-7316, USA}

\date{\today}
\begin{abstract}
In this paper, we study the magnetic reconnection process of energy extraction from a rapidly rotating Kerr-Newman-modified-gravity (MOG) black hole by investigating the combined effect of black hole charge and the MOG parameter. We explore the energy efficiency of energy extraction and power by applying the new energy extraction mechanism proposed by Comisso and Asenjo. Based on an attractive gravitational charge of the MOG parameter $\alpha$ that physically manifests to strengthen black hole gravity, we show that the combined effect of the MOG parameter and black hole charge can play an increasingly important role and accordingly lead to high energy efficiency and power for the energy extraction via the magnetic reconnection. Further, we study to estimate the rate of energy extraction under the fast magnetic reconnection by comparing the power of the magnetic reconnection and Blandford-Znajek (BZ) mechanisms. We show that the rate of energy extraction increases as a consequence of the combined effect of black hole charge and MOG parameter. It suggests that magnetic reconnection is significantly more efficient than the BZ mechanism. In fact, the magnetic reconnection is fueled by magnetic field energy due to the twisting of magnetic field lines around the black hole for the plasma acceleration, and thus the MOG parameter gives rise to even more fast spin that can strongly change the magnetic field reconfiguration due to the frame dragging effect. This is how energy extraction is strongly enhanced through the magnetic reconnection, thus making the energy extraction surprisingly more efficient for the Kerr-Newman-MOG black hole than the Kerr black hole under the combined effect of black hole charge and MOG parameter.

\end{abstract}
\maketitle

\section{Introduction}
\label{introduction}

In general relativity (GR) astrophysical black holes are formed as a consequence of the end state of evolution of massive stars via gravitational collapse and are regarded as the most intriguing and fascinating objects not only for their extreme geometric and
remarkable gravitational aspects but also for their important role in explaining highly energetic astrophysical events. In astrophysical phenomena such as active galactic nuclei (AGN)  \cite{Peterson:97book}, $\gamma$-ray bursts \cite{Meszaros06}, and ultraluminous x-ray binaries \cite{King01ApJ} an enormous amount of energy is released. It is believed that astrophysical black holes come into play a key role in these extremely powerful astrophysical phenomena \cite{McHardy06,Woosley93ApJ,Preparata98,Popham98,Rappaport05}. Black holes can therefore be considered the most powerful energy sources in the Universe and are related to astronomical observations associated with outflows from active galactic nuclei with the energy, which is of order $10^{42}$-$10^{47}\:\rm{erg/s}$ ~\cite{Fender04mnrs,Auchettl17ApJ,IceCube17b}.  However, the tremendous amount of energy produced by these events is believed to have two likely origins: i) It might be the gravitational potential energy of matter falling towards the black hole during the accretion phase or it might be the energy of falling matter during gravitational collapse phase of a black hole. ii) Second, it might possibly be the black hole's own energy, as predicted by general relativity in the case of a rotating black hole. As a consequence, understanding the genesis of the highly powerful astrophysical phenomena in the vicinity of a black hole has profound ramifications and is fundamentally important. 

The question, could rotational energy be extracted out from a rotating black hole was first formulated by Penrose \cite{Penrose:1969pc} and it was shown that it can be extracted. What happens is the particle that comes from infinity could have negative energy with respect to infinity and splits into two parts in the ergosphere, i.e., one falls into a black hole with negative energy and the other escapes with energy greater than that of the original particle. This is how black hole rotational energy $E_{rot}=\left(1-1/\sqrt{2}\right)M\approx 0.29 M$ could be extracted out via the Penrose process. Later, this process was then extended to many various situations; here we give some references. The effect of gravitomagnetic charge of the black hole on the energy that can be extracted by the Penrose process was addressed \cite{Abdujabbarov11} and it was found that its impact increases the amount of the extracted energy. The Penrose process was also extended to the spinning test particles~\cite{Okabayashi20} and rotating regular black hole~\cite{Toshmatov:2014qja}. 

On the other hand, the magnetic field becomes increasingly important in modeling new alternative thought mechanisms for extracting out rotational energy from rotating black holes. In this respect, Blandford and Znajek first addressed the effect of the magnetic field on the energy extraction process in the accretion disks of AGN~\cite{McKinney07} and it has since been known as the Blandford-Znajek (BZ) mechanism~\cite{Blandford1977}. The impact of the magnetic field has been extended to a large variety of situations over the years~\cite{Wagh89,Morozova14,Alic12ApJ,Moesta12ApJ,ALLOQULOV2023302}. Later, to bring out the impact of magnetic field on the efficiency of energy extraction the Pensose process was considered as the magnetic Penrose process~\cite{Bhat85,Parthasarathy86}. It has since been shown that the magnetic Penrose process becomes a more effective process for extracting the energy from rotating black holes through the magnetic field~\cite{Nozawa05,Dadhich18mnras,Shaymatov22b,Kurbonov2023}. 

Further, a thought experiment for these highly powerful astrophysical phenomena was proposed by Ba$\mathrm{\tilde{n}}$ados \textit{et} \textit{al.} by considering high energy particle collisions in the close vicinity of the black hole horizon \cite{banados09} and it was shown that it leads to high energy that can be extracted through this process. For that an extremal Kerr black hole can act as a particle accelerator to arbitrary high energies produced by the collision of two particles. Following Ba$\mathrm{\tilde{n}}$ados \textit{et} \textit{al.}~\cite{banados09} there has since been an extensive body of work considering various contexts
~\cite{Grib11,Jacobson10,Harada11b,Wei10,Zaslavskii10,Kimura11,Banados11a,Frolov12,Liu11,Igata12,Shaymatov13,Tursunov13,Shaymatov18a,Shaymatov21pdu,Turimov23PLB,Turimov2023}.

A new thought explanation for high energy observations has recently been proposed independently by considering the magnetic reconnection process near the horizon of a rotating black hole. It is envisaged by the fact that the frame dragging effect of a rotating black hole can twist the magnetic field lines, thus resulting in causing antiparallel magnetic field lines in the equatorial plane. The magnetic reconnection process occurring in the ergosphere on the surrounding environment of a black hole can accelerate particles, some of which do, however, attain negative energy and get absorbed by the black hole while the other accelerated particles come out with positive energy that can be extracted from the black hole as the stolen energy. This is how rotational energy of the black hole could be driven out through the magnetic reconnection occurring continuously inside the ergosphere because of the black hole's fast spin. This defines the main difference in contrast to the above mentioned mechanisms for energy extraction. Analysis associated with this process was addressed by Koide and Arai \cite{Koide08ApJ} using the slow magnetic reconnection so as to attempt the energy extraction from black hole. It was then realized that this scenario would not be viable to extract out the energy from the black hole. It was, however, suggested that the relativistic reconnection needs to be required as the most promising condition for energy extraction from a black hole as a form of the outflow jets due to the magnetic field configuration. Later, it was also suggested that the accelerated particles under the magnetic reconnection can attain negative energy as stated by the general-relativistic kinetic simulations and the energy extraction that could be driven out from the black hole due to these negative energy particles would be comparable to the one extracted by the BZ mechanism~\cite{Parfrey19PRL}. However, there existed no evaluation of energy released through magnetic reconnection.

Recently, unlike the above mentioned scenarios, the energy extraction through the magnetic reconnection process was approached from a different perspective by Comisso and Asenjo \cite{Comisso21} considering a rapidly spinning Kerr black hole. To that, a novel mechanism was first proposed by Comisso and Asenjo, thus allowing one to compute the energy efficiency and the power for energy extraction via the magnetic reconnection process. It was shown that this novel mechanism can be considered as an efficient mechanism for energy extraction since the efficiency and the power get affected strongly by black hole spin through the magnetic reconnection. It is also worth noting that this mechanism is now a well accepted mechanism for energy extraction via the magnetic reconnection, accordingly referred to as the Comisso-Asenjo mechanism which we further apply for our detailed analysis in this study. The relevance of the energy extraction from the Comisso-Asenjo mechanism via the magnetic reconnection has since been considered in several recent investigations for rapidly spinning black holes \cite{Liu22ApJ,Wei22,Carleo22,Khodadi22,Wang22}. {We also noticed that the Comisso-Asenjo mechanism has been considered for a rotating MOG black hole~\cite{Khodadi_2023} addressing the impact of MOG field on the energy extraction process. We do however emphasize that our approach can generalize all, thus bringing out the combined effect of black hole charge and the MOG parameter on the energy extraction via the magnetic reconnection process and providing the interpretation of the MOG parameter as an attractive gravitational charge. }

In this paper, we consider a rotating Kerr-Newman-MOG black hole immersed in an external magnetic field, as presented by the line element described in Ref.~\cite{Moffat15MOG}. For this black hole spacetime geometry, we study the energy extraction from a rotating Kerr-Newman-MOG black hole and analyze the impact of this geometry on the magnetic reconnection process from evaluating the energy efficiency of energy extraction and the power, which are given as a function of black hole spin and MOG parameter, the location, plasma magnetization parameter, and magnetic field orientation by imposing all required conditions. We also explore the parameter space of black hole parameters, ergoregion and the innermost stable circular orbit (ISCO) for charged test particles.  

This paper is organized as follows: In Sec.~\ref{Sec:KN-MOG} we briefly describe the Kerr-Newman-MOG black hole spacetime and particle dynamics. In Sec.~\ref{Sec:MR} we explore the impact of the rotating Kerr-Newman-MOG black hole (BH) on the magnetic reconnection and further study the energy extraction by the magnetic reconnection. We further explore the power, the energy efficiency, and the rate of energy extraction through the magnetic reconnection mechanism in Sec.~\ref{Sec:power-mm}. Finally, we end up with concluding remarks in Sec.~\ref{Sec:conclusion}. {Throughout the manuscript we use a system of units in which $G_{\rm{N}}=c=1$.}

\section{Kerr-Newman-MOG black hole metric and particle dynamics }\label{Sec:KN-MOG}

In Boyer-Lindquist coordinates, the background geometry of the Kerr-Newman-MOG spacetime is given by \cite{Moffat15MOG}
\begin{eqnarray}
ds^2 &=& -\frac{\Delta}{\rho^2} [dt-a\sin^2 \theta d\phi ]^2 +\rho^2 \left[ \frac{dr^2}{\Delta} + d\theta^2 \right] \nonumber \\
&+&\frac{\sin^2 \theta}{\rho^2} \left[ (r^2+a^2)d\phi -adt \right]^2\ , 
\end{eqnarray}
where
\begin{eqnarray}
\rho^2 &=& r^2 + a^2 \cos^2 \theta\ , \nonumber \\
\Delta &=& r^2-2G_{\rm{N}}(1+\alpha)Mr + a^2 +Q^2\nonumber\\&& +\, G^2_{\rm{N}}\alpha (1+\alpha) M^2\ . 
\end{eqnarray}
The MOG parameter $\alpha$ is a dimensionless measure of the difference between the Newtonian gravitational constant $G_{\rm{N}}$ and the additional gravitational constant $G$
\begin{equation}
\alpha=\frac{G-G_{\rm{N}}}{G_{\rm{N}}}\ .
\end{equation}
The Arnowitt-Deser-Misner (ADM) mass and the angular momentum of the Kerr-MOG black hole are given by \cite{Moffat15MOG,Sheoran18}
\begin{eqnarray}
\mathcal{M}=(1+ \alpha)M\, .
\end{eqnarray}
\begin{figure}
    \centering
    \includegraphics[scale=0.7]{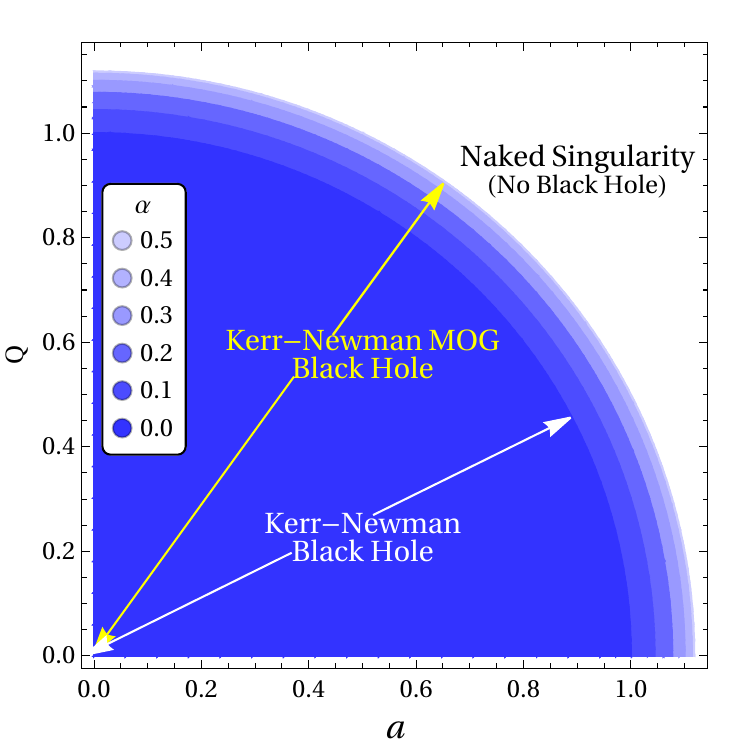}
    \caption{%\cred
    {The parameter space plot between the charge parameter $Q$ and the spin parameter $a$ of the Kerr-Newman-MOG BH for the MOG parameter $\alpha$ in the range $0$-$0.5$. 
    %and $0.5$ to $1.0$, respectively. Bottom panel shows parameter space plot between the MOG parameter $\alpha$ and the spin parameter $a$ of the Kerr-Newman-MOG BH for charge parameter $Q$ in the range $0$ to $0.95$. Here, the shaded curves separate the BH from no BH regions (i.e., where, there is no real root of $\Delta= 0$). 
    Additionally, the parameter $M$ is set to unity throughout.}}
    \label{fig:Region_Plot_btw_alpha_n_a}
\end{figure}
\begin{figure}
    \centering
    \includegraphics[scale=0.55]{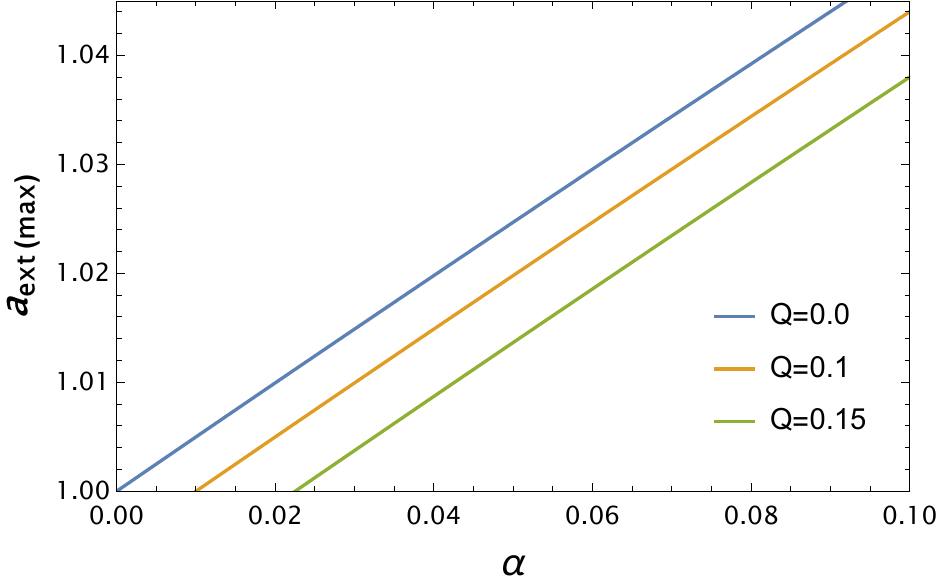}
    \caption{%\cred
    {An extremal (the maximal) value of black hole spin parameter against the MOG parameter $\alpha$ for various combinations of the charge parameter $Q$. }}
    \label{fig:aext}
\end{figure}

The function $\Delta$ can be rewritten in terms of the ADM mass,
\begin{equation}
\Delta=r^2 -2\mathcal{M}r+a^2+Q^2+\frac{\alpha}{1+\alpha}\mathcal{M}^2 \label{delta}\ , 
\end{equation}
where we have set $G_{\rm{N}}=1$ without loss of generality. The spatial locations of the horizons are the roots of $\Delta$ as 
\begin{equation}
r_{H}=\mathcal{M} \pm \sqrt{\frac{\mathcal{M}^2}{1+\alpha}-a^2-Q^2} \ . \label{rplus}
\end{equation}
Notice that the parameters of the Kerr-Newman-MOG spacetime represent a black hole surrounded by an event horizon provided that
\begin{equation} 
\mathcal{M}^2 \geq (1+\alpha) \left(a^2+Q^2\right)\, ,\label{criterion}
\end{equation}
%\cred
{where the inequality corresponds to the case of an extremal black hole, i.e., $a\to a_{ext}$ that refers to the maximal spin of the black hole. If it is not satisfied the black hole can no longer exist. On the other hand, there also exists another key point that the Kerr-Newman-MOG black hole rotates with the spin greater than that of one for the Kerr black hole due to the effect of MOG parameter $\alpha$; see in Fig.~\ref{fig:Region_Plot_btw_alpha_n_a}. In Fig.~\ref{fig:Region_Plot_btw_alpha_n_a}, we demonstrate the parameter space plot between the charge parameter $Q$ and the spin parameter $a$ of the Kerr-Newman-MOG BH for various combinations of the MOG parameter $\alpha$ in the range $0$-$0.5$, respectively. It is interestingly seen from Fig.~\ref{fig:Region_Plot_btw_alpha_n_a} that the radius of parameter space increases as the MOG parameter increases. As can be seen from Fig.~\ref{fig:Region_Plot_btw_alpha_n_a}, the black hole can exist in the shaded region which is separated from no black hole regions by the curves. We note that $\Delta= 0$ always has a real root for any value of the spin $a$ and charge $Q$ parameter in the case of MOG parameter $\alpha$. Also, in Fig.~\ref{fig:aext} we demonstrate the dependence of an extremal value of the spin parameter of the Kerr-Newman-MOG BH on the MOG parameter $\alpha$ for various combinations of the charge parameter $Q$. As is explicitly shown by Fig.~\ref{fig:aext}, $a_{ext}$ increases as the MOG parameter $\alpha$ increases, while its curves shift downward toward down to its smaller values as a consequence of an increase in the value of the charge parameter $Q$. }

%In top and middle panels of Fig.~\ref{fig:Region_Plot_btw_alpha_n_a}, we demonstrate the parameter space plot between the charge parameter $Q$ and the spin parameter $a$ of the Kerr-Newman-MOG BH for various combinations of the MOG parameter $\alpha$ in the range $0$ to $0.5$ and $0.5$ to $1.0$, respectively. It is interestingly seen from Fig.~\ref{fig:Region_Plot_btw_alpha_n_a} that the radius of parameter space increases as the MOG parameter increases up to  $\alpha=0.5$. It does however decrease for $0.5<\alpha<1$. Similarly, in bottom panel of Fig.~\ref{fig:Region_Plot_btw_alpha_n_a}  we show parameter space plot between the MOG parameter $\alpha$ and the spin parameter $a$ of the Kerr-Newman-MOG BH for various combinations of black hole charge parameter $Q$ in the range $0$ to $0.95$. As can be seen from Fig.~\ref{fig:Region_Plot_btw_alpha_n_a} black hole can exist in the shaded region which is separated from no black hole regions by the curves. We note that $\Delta= 0$ always has real root for any value of the spin $a$ and charge $Q$ parameter in the case of MOG parameter $\alpha$. 

\begin{figure*}
    \centering
    \includegraphics[scale=0.4]{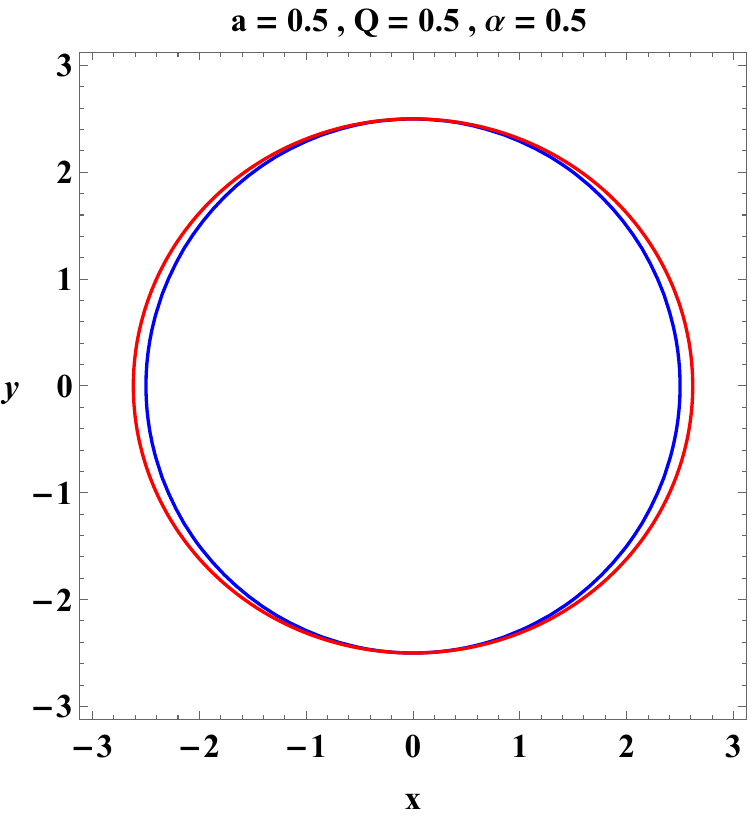}
    \includegraphics[scale=0.4]{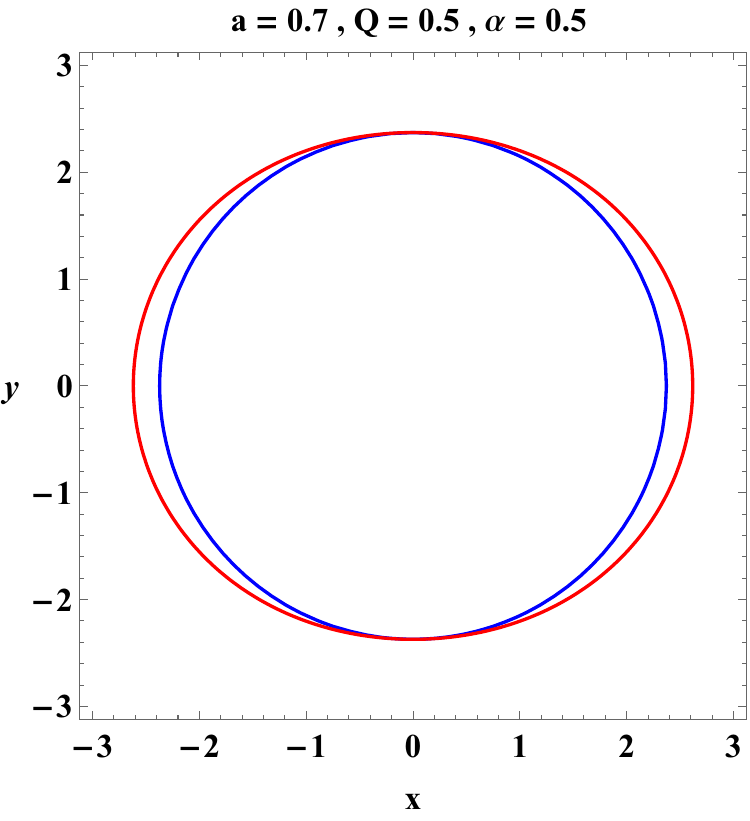}
    \includegraphics[scale=0.4]{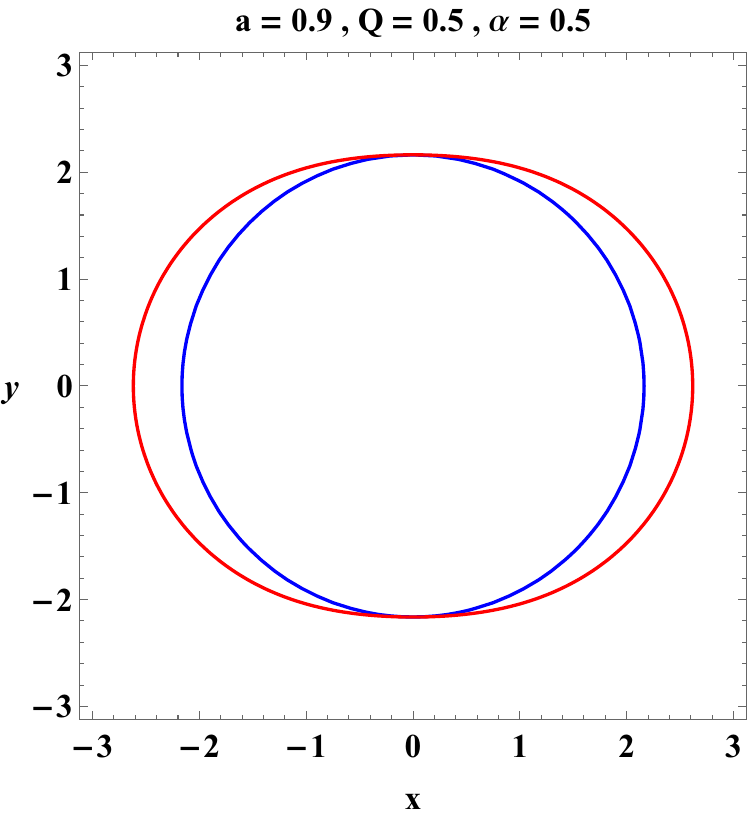}
     \includegraphics[scale=0.4]{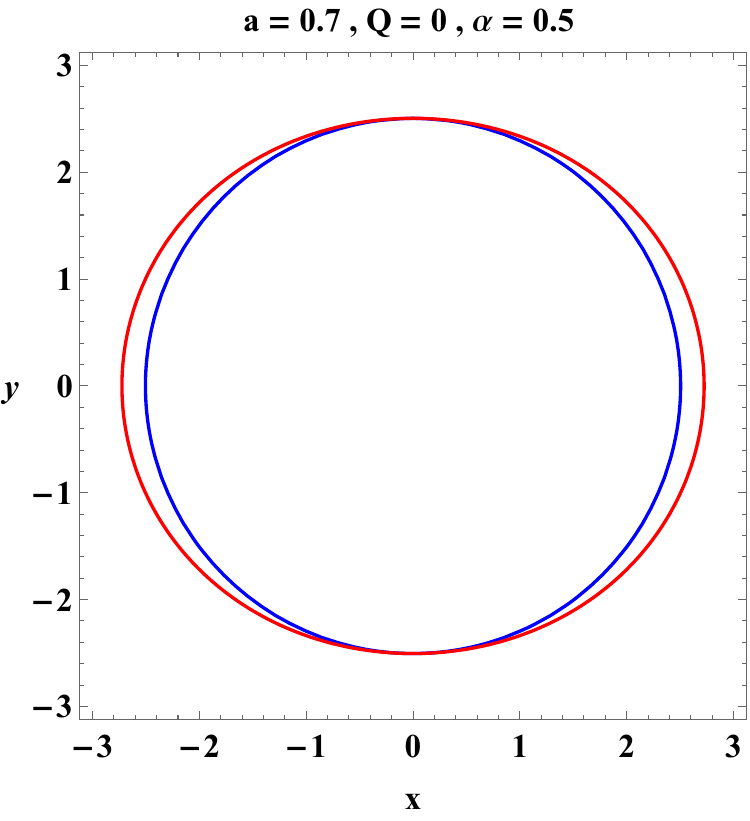}
    \includegraphics[scale=0.4]{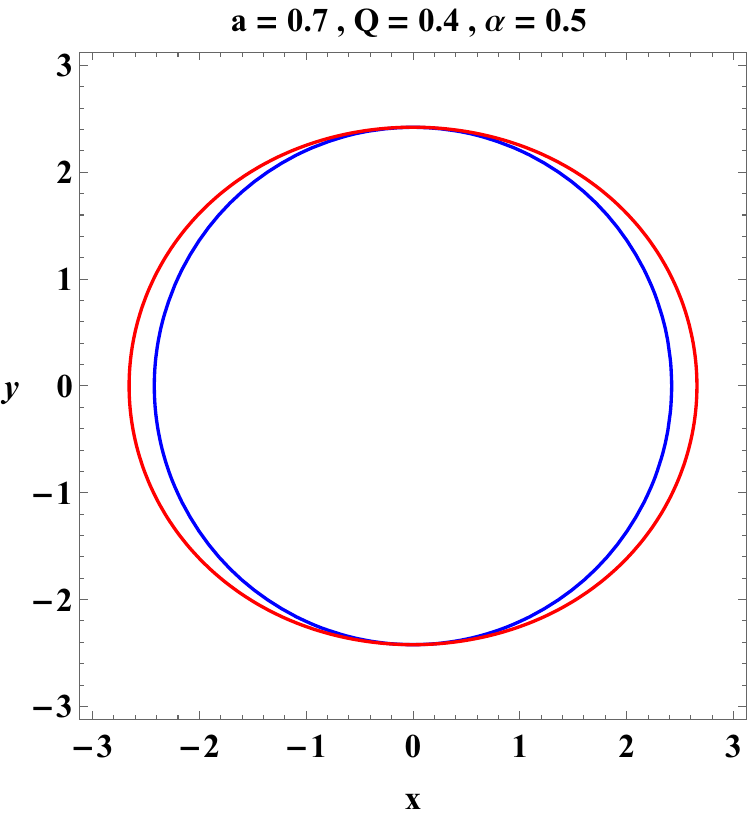}
    \includegraphics[scale=0.4]{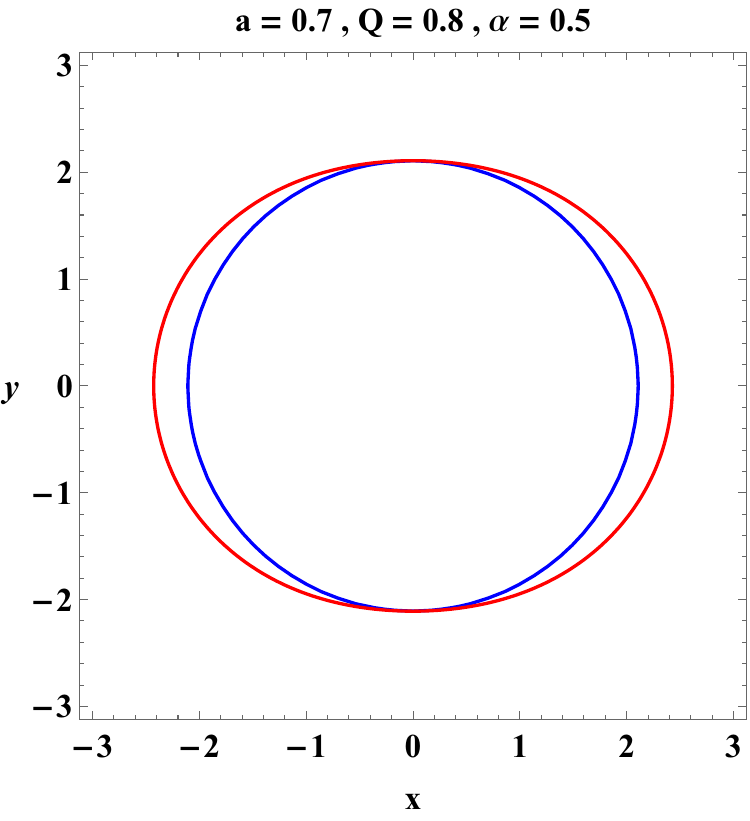}
     \includegraphics[scale=0.4]{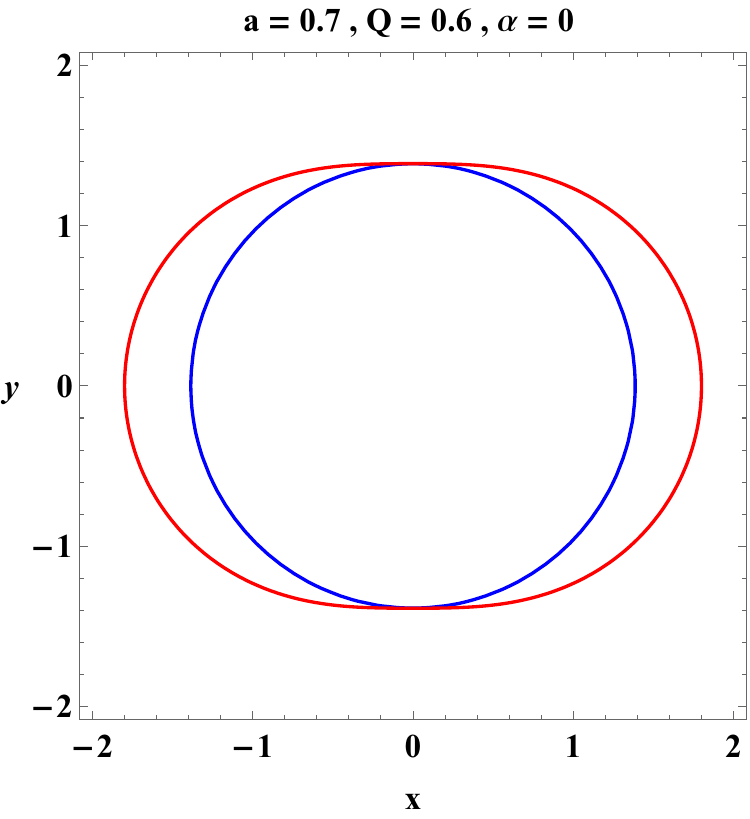}
    \includegraphics[scale=0.4]{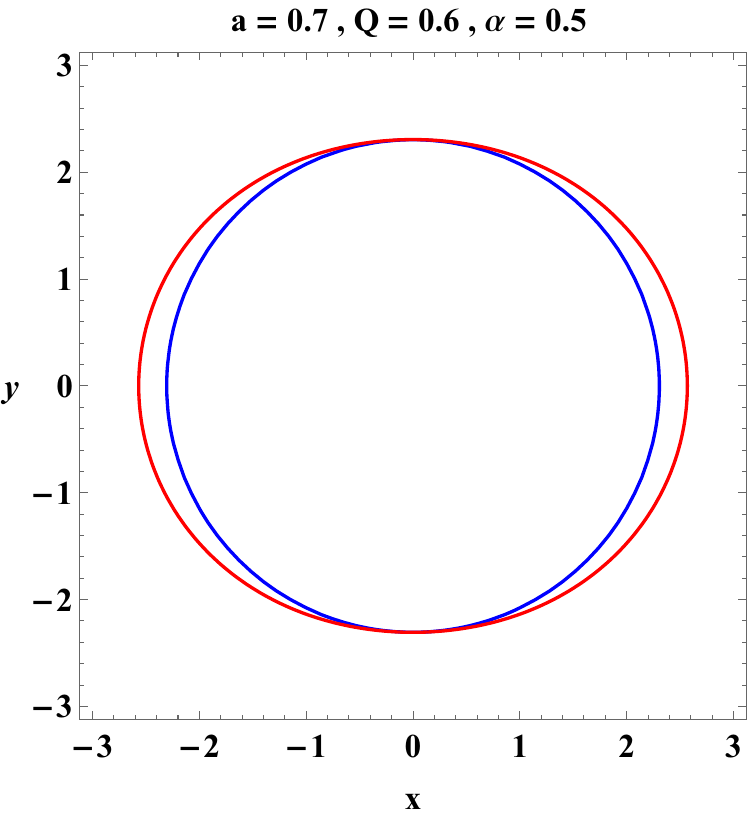}
    \includegraphics[scale=0.4]{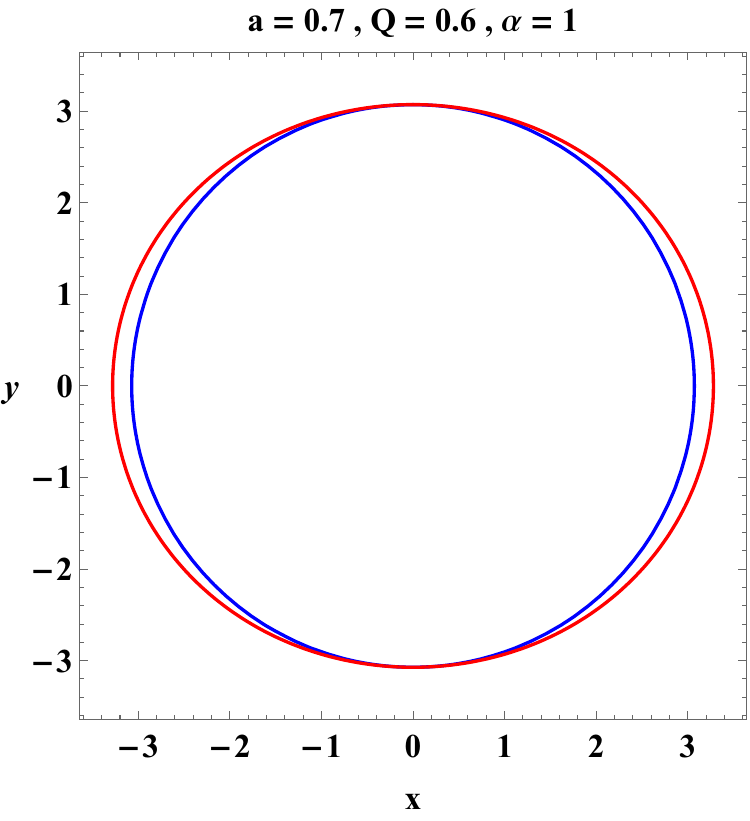}
     \caption{Ergospehere around the Kerr-Newman-MOG black hole for various possible cases. Top row: ergospehere is plotted for various combinations of spin parameter $a$ for fixed $Q$ and $\alpha$. Middle row: ergospehere is plotted for various combinations of black hole charge $Q$ for fixed $a$ and $\alpha$. Bottom row: ergospehere is plotted for various combinations of MOG parameter $\alpha$ for fixed $a$ and $Q$. Note that red and blue lines respectively correspond to the radius of ergosphere and black hole horizon. }
    \label{fig:ergo}
\end{figure*}

There is also another static surface that is estimated by the timelike Killing vector $\xi^{\mu}_{(t)} = \partial/\partial t$, i.e., $g_{tt}= 0$ which solves to give 
\begin{eqnarray}\label{Eq:Ergo}
r_{E}=\mathcal{M} +\sqrt{\frac{\mathcal{M}^2}{1+\alpha}-Q^2-a^2\cos^2\theta}\, .
\end{eqnarray}
Note that it is not possible for any particle to be static at a fixed point below this surface. Thus, the region existing between the surface $r_{E}$ and the horizon $r_{H}$ refers to the ergosphere, where a timelike particle's energy $E$ may become negative relative to an observer at infinity. In Fig.~\ref{fig:ergo} we plot the behavior of the ergosphere for various possible cases in the presence of black hole charge and MOG parameters.   
{Following to Wald \cite{Wald74} we consider the generalized form of vector potential of the electromagnetic field around the Kerr-Newman MOG black hole, which is given by the following form \cite{Mustapha16}: } 
%
%\textcolor{red}{\begin{eqnarray}\label{Eq:vec-pot}A^{\mu}&=&\Big[\frac{-Qr}{r^2-2f}+aB\Big(1+\frac{f_2 \sin^2{\theta}}{r^2}\Big)\Big]\xi^{\mu}_{t} \nonumber\\&+&\Big[\frac{B}{2}\Big(1+\frac{2f_2}{r^2}\Big)-\frac{Qa}{r(r^2-2f)}\Big]\xi^{\mu}_{\phi},\end{eqnarray}}
%
\begin{eqnarray}\label{Eq:vec-pot}
A^{\mu}&=&\Big[\frac{Qr}{r^2-2f}-aB\Big(1+\frac{f_2 \sin^2{\theta}}{r^2}\Big)\Big]\xi^{\mu}_{t} \nonumber\\
&+&\Big[-\frac{B}{2}\Big(1+\frac{2f_2}{r^2}\Big)+\frac{Qa}{r(r^2-2f)}\Big]\xi^{\mu}_{\phi},
\end{eqnarray}
where we have defined $f=f_1 r+f_2$. Note that $f_1$ and $f_2$ are defined by
\begin{eqnarray}
f_1&=&(1+\alpha)M\, ,\\ 
f_2&=& -(1+\alpha)\left(\alpha M^2+Q^2\right)/2\, .
\end{eqnarray}

Let us then consider a particle motion with rest mass $m$ and charge $q$ around the Kerr-Newman-MOG black hole immersed in an external magnetic field. Note that particle motion at the circular orbit around the black hole is also an important point for the magnetic reconnection process which we further explore in this paper. The magnetic field can affect the charged particle motion drastically due to the strong Lorentz force even though it is small enough \cite{Frolov12,Shaymatov18a}. From the asymptotic properties of the black hole the magnetic field is assumed to be uniform and oriented along the axis of symmetry of the black hole. The Hamiltonian that completely describes the system can be written as follows \cite{Misner73}: 
\begin{equation}
   H \equiv \frac{1}{2}g^{\alpha\beta}(\pi_\alpha - q A_\alpha)(\pi_\beta - q A_\beta)\, , \label{eq:hamiltonian}
\end{equation}
where we have defined $\pi_\alpha$, which is referred to as the canonical momentum of a charged test particle. Here, $A_\alpha=A_\mu(r,\theta)$ refers to the four-vector potential of the electromagnetic field and is defined by Eq.~(\ref{Eq:vec-pot}), while the Hamiltonian turns out to be constant as $H=-m^2/2$.

\begin{table*}
\begin{center}
\caption{Numerical values of the $\mathcal{L}_{ISCO}, \mathcal{E}_{ISCO}, r_{ISCO}$ for the test particle moving at the ISCO radius around Kerr-Newman-MOG black hole for different possible cases.  Note that we restrict ourselves to the equatorial plane, $\theta=\pi/2$, and to  the case in which we set $a/M=0.2$, $Q/M=0.6$, and $BM=0.01 $ in order to bring out the effect of the MOG parameter on the ISCO parameters.}\label{table1}
\resizebox{.7\textwidth}{!}
{
\begin{tabular}{l l|l l l | l l l }
% \begin{tabular}{c c c c | c c | c c | c c}
 \hline \hline
 & & &\multicolumn{1}{c}{}  $\alpha=0.1$  &\multicolumn{1}{c}{} & &$\alpha=0.3$  & %&\multicolumn{2}{c}{} & &
 \\
\cline{3-5}\cline{6-8}
& $q>0$    & $\mathcal{L}_{ISCO}$ & $\mathcal{E}_{ISCO}$ & $r_{ISCO}$    &$\mathcal{L}_{ISCO}$ & $\mathcal{E}_{ISCO}$ & $r_{ISCO}$   \\
\hline
& 0.1   & 3.44637 & 0.922775  & 5.19946  & 4.05231 & 0.923171  &6.13550   \\
& 0.3   & 3.61588 & 0.912702  & 5.22805  & 4.21817 & 0.913581  &6.15809 \\
& 0.5     & 3.78114 & 0.902597 & 5.26119  & 4.38026 & 0.903916 & 6.18443  \\
& 0.7       & 3.94263 & 0.892462 & 5.29771  & 4.53895 & 0.894182 & 6.21365  \\
& 0.9       & 4.10072 & 0.882296 & 5.33676  & 4.69454 & 0.884382 & 6.24509  \\
\hline
& & &\multicolumn{1}{c}{} $\alpha=0.1$  &\multicolumn{1}{c}{} & 
 &$\alpha=0.3$ &  %&\multicolumn{2}{c}{}   
\\
\cline{3-5}\cline{6-8}
& $q<0$    & $\mathcal{L}_{ISCO}$ & $\mathcal{E}_{ISCO}$ & $r_{ISCO}$ & $\mathcal{L}_{ISCO}$ & $\mathcal{E}_{ISCO}$ & $r_{ISCO}$ \\  
\hline
& $-$0.1   & 3.42571 & 0.932808 & 5.41592  & 3.88225 & 0.932680 & 6.11777 \\
& $-$0.3   & 3.24710 & 0.942622 & 5.40325 & 3.70744 & 0.942099 & 6.10632  \\
& $-$0.5     & 3.06222 & 0.952364 & 5.40154  & 3.52722 & 0.951417 & 6.10308 \\
& $-$0.7       & 2.86984 & 0.962015 & 5.41515  & 3.34071 & 0.960618 & 6.11062  \\
& $-$0.9       & 2.66824 & 0.971543 & 5.45088  & 3.14675 & 0.969680 & 6.13251  \\
 \hline \hline
\end{tabular}
}
\end{center}
\end{table*}

For the charged particle, its four-momentum can be written as 
\begin{equation}\label{eq:h-0}
   p^\alpha \equiv \frac{dx^\alpha}{d\lambda} = g^{\alpha\beta}(\pi_\beta - q A_\beta)\, ,
\end{equation}
where $\lambda=\tau/m$ corresponds to an affine parameter with the proper time $\tau$. Let us then write Hamilton's equations of motion in terms of $x^\alpha$ and $\pi_\alpha$, which are given by
\begin{eqnarray}
  \frac{dx^\alpha}{d\lambda} = \frac{\partial H}{\partial \pi_\alpha}\,\mbox{~~and~~} 
%\label{eq:h-1} 
  \frac{d\pi_\alpha}{d\lambda} = - \frac{\partial H}{\partial x^\alpha}\, .\label{eq:h-2}
\end{eqnarray}
We note that we further restrict our attention to the equatorial plane, i.e., $\theta=\pi/2$. As was mentioned, the first one in the above equations is referred to as a constraint equation that defines the four-momentum of the charged particle. Thus, the equations of motion for the charged particle can be defined as
\begin{eqnarray}
\label{tt} p^{t}&=&\frac{1}{r^2}\left[ a \big(\pi_{\varphi}+a\pi_{t}\big)+\frac{r^2+a^2}{\Delta}P\right], \\ \nonumber \\
\label{ff}
p^{\varphi}&=&\frac{1}{r^2}\left[ \big(\pi_{\varphi}+a\pi_{t}\big)+\frac{a}{\Delta}P\right] ,\\ \nonumber\\
\label{rr}
p^{r}&=&\left(\frac{P^2-\Delta\big[r^2+\big(\pi_{\varphi}+a\pi_{t}\big)^2\big]}{r^4}\right)^{1/2}\, ,
\end{eqnarray}
where we have defined $P=(r^2+a^2)(-\pi_{t})-a \pi_{\varphi}$. For the Hamilton-Jacobi equation, the action $S$ can be separated in such a way that it takes the form as 
\begin{eqnarray}\label{separation}
S= \frac{1}{2}m^2\lambda-Et+L\varphi+S_{r}(r)+S_{\theta}(\theta)\, .
\end{eqnarray}
Here, accordingly $E \equiv -\pi_t$ and $L \equiv \pi_{\varphi}$ are constants of motion and correspond to the specific energy and angular momentum of the charged particle, respectively. There exists a fourth constant of motion except the rest mass of the test particle $m$, which is related to the latitudinal motion. However, we further omit this constant since we focus on the motion that takes place in the equatorial plane.
The rest terms such as $S_{r}$ and $S_{\theta}$ turn out to be functions related to $r$ and $\theta$, respectively. Following Eqs.~(\ref{eq:hamiltonian}) and (\ref{separation}), the Hamilton-Jacobi equation is given by 
\begin{eqnarray}\label{Eq:separable}
&-&\left[\frac{(r^2+a^2)^2}{\Delta}-a^2\sin^2\theta\right](E+qA_{t})^2
+\Delta
\left(\frac{\partial S_{r}}{\partial r}\right)^2\nonumber\\
&+&\frac{2a \left(2\mathcal{M}-\frac{\alpha}{r(1+\alpha)}\mathcal{M}^2-\frac{Q^2}{r}\right)r}{\Delta}(E+qA_{t})(L-qA_{\varphi})\nonumber\\&+&\left(\frac{\partial S_{\theta}}{\partial \theta}\right)^2+
\left[\frac{1}{\sin^2\theta}-\frac{a^2}{\Delta}\right](L-qA_{\varphi})^2+m^2\Sigma=0\, .\nonumber\\
\end{eqnarray}
On the basis of Eq.~(\ref{Eq:separable}), the radial equation of motion for the charged particle moving on the equatorial plane (i.e. $\theta=\pi/2=const$) can be defined by 
\begin{eqnarray}
\frac{1}{2}\dot{r}^{2} + V_{eff}(r)=0,
\end{eqnarray}
where $V_{eff}(r)$ refers to the effective potential for radial motion and is given in general form as 
%\begin{eqnarray*}
%V_{eff}(r,\mathcal{E},\mathcal{L}) &=&...\, \mbox{\textcolor{blue}{write its form right here}}\, . \end{eqnarray*}
\begin{eqnarray}\label{Veff1}
   V_{eff}(r)&=&-\frac{1}{2r^2}\left[\left(r^2+a^2+\frac{2A(r)a^2}{r}\right)\right.%\nonumber\\&\times &
\left(\mathcal{E}+\frac{q}{m}A_{t} 
\right)^2 \nonumber\\&-&\left(1-\frac{2A(r)}{r}\right)\left(\mathcal{L}-\frac{q}{m}A_{\phi}\right)^2-\Delta \nonumber\\&-&\left.\frac{4A(r)\,a\,\left(\mathcal{E}+q/mA_{t}\right)\,\left(\mathcal{L}-q/mA_{\phi}\right)}{r}\right]\, , 
\end{eqnarray}
with 
\begin{eqnarray*}
A(r)=\left(\mathcal{M}-\frac{\alpha}{2\,r(1+\alpha)}\mathcal{M}^2-\frac{Q^2}{2r}\right)\, .
\end{eqnarray*}
Here we have defined specific constants as $\mathcal{E}=E/m$, $\mathcal{L}=L/m$. 

We then turn to the innermost stable circular orbit for test particles moving around the black hole as the ISCO radius is an important key for the magnetic reconnection process that occurs in the ergo region. To determine the ISCO, the following standard conditions must be satisfied: 
\begin{eqnarray}\label{Eq:circular}
V_{eff}(r)=0, \mbox{~~~}  V_{eff}^{\prime}(r)=0\, \mbox{~~and~~} V_{eff}^{\prime\prime}(r)\geq 0\, ,
\end{eqnarray}
where we have defined $^{\prime}$ as a derivative with respect to $r$.  We here note that since the analytic forms of the ISCO turns to be very long and complicated expressions for explicit display we shall for simplicity further resort to numerical evaluation for further analysis. For magnetic reconnection, one needs to consider corotating orbits as it occurs in the ergosphere near the black hole horizon.  In Table \ref{table1}, we demonstrate the numerical values of the ISCO parameters for the test particle moving at the ISCO radius around the Kerr-Newman-MOG black hole for various possible cases. Here we mainly focus on the MOG parameter to understand more deeply its impact on the ISCO parameters, as seen in Table~\ref{table1}. We, therefore, consider black hole spin and charge parameters to be small. As can be seen from Table~\ref{table1}, the ISCO radius grows as a consequence of an increase in the value of the MOG parameter. One can then deduce that this is consistent with the interpretation of the MOG parameter as an attractive gravitational charge that physically manifests to strengthen black hole gravity. Taking this remarkable aspect of the MOG parameter $\alpha$  into consideration plays a very crucial role in modeling the magnetic reconnection process for the Kerr-Newman-MOG black hole. We do therefore consider this key point to evaluate the efficiency of energy extraction, power, the phase-space region, and the rate of energy extraction via magnetic reconnection for further analysis.

\section{Energy extraction through magnetic reconnection 
process  }\label{Sec:MR}

\begin{figure*} 
\begin{center} \begin{tabular}{c c} \includegraphics[scale=0.5]{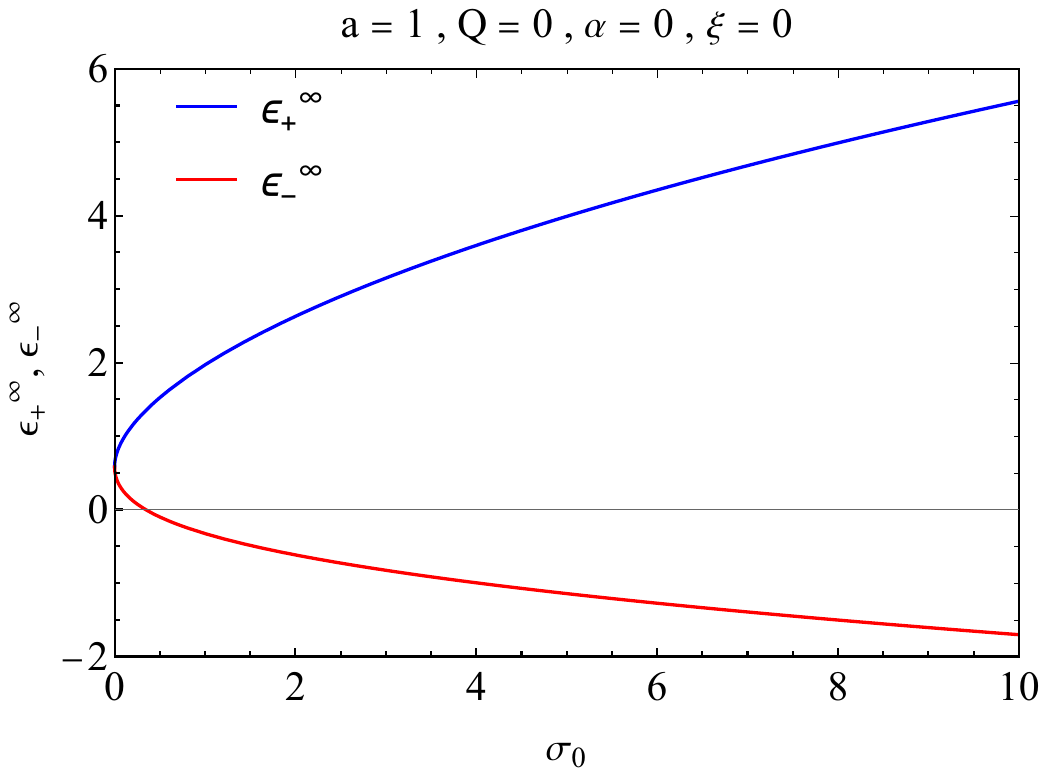}\hspace{-0.2cm} \includegraphics[scale=0.385]{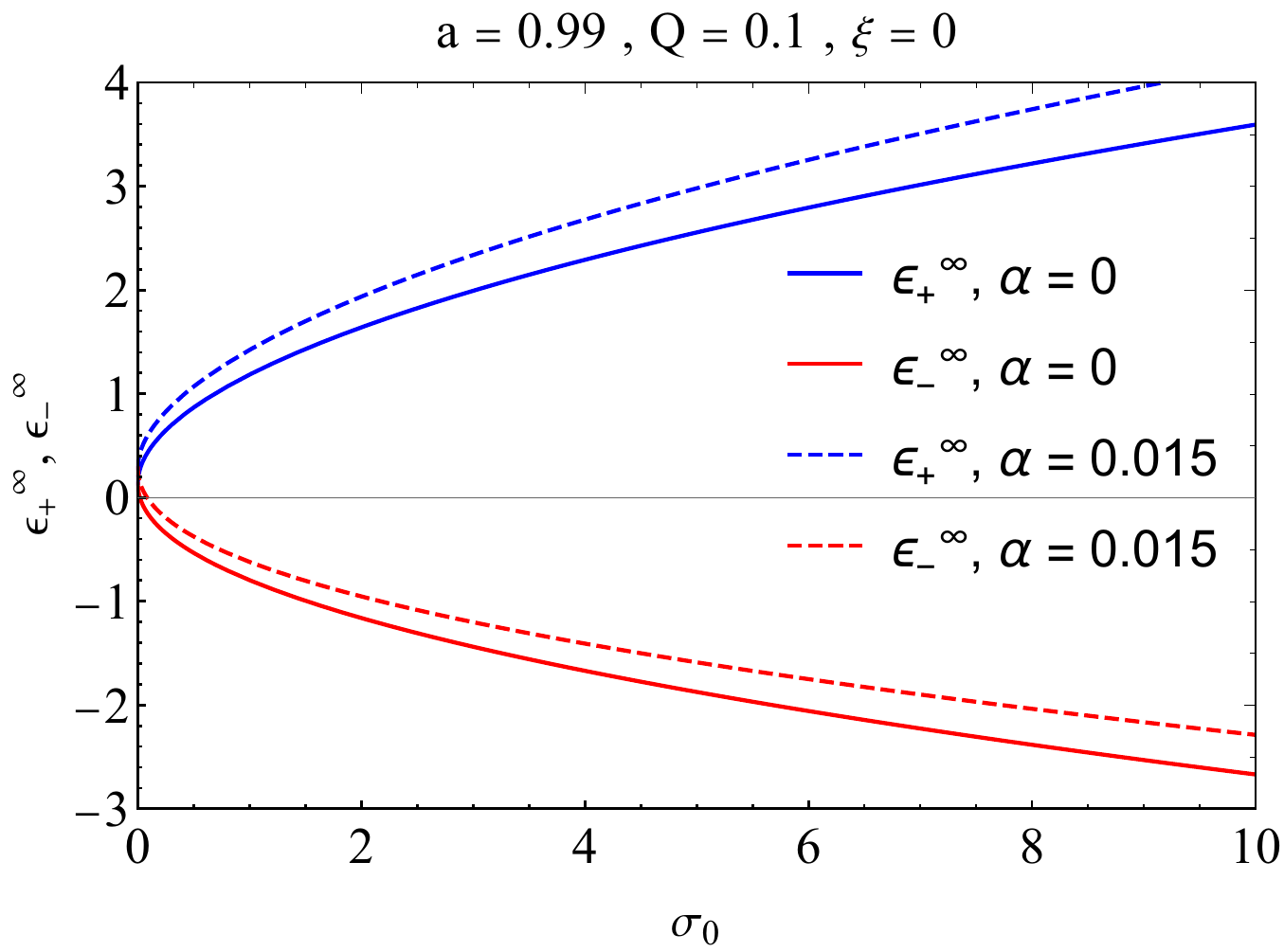}\hspace{0cm} 
\end{tabular} \caption{\label{Fig:energy} {The energies $\epsilon_{+}$ and $\epsilon_{-}$ per enthalpy at infinity are plotted for various possible cases. Left: $\epsilon_{+}$ and $\epsilon_{-}$ are plotted for $Q=0$ and $\alpha=0$. Right: $\epsilon_{+}$ and $\epsilon_{-}$ are plotted as a consequence of the presence of the MOG parameter while keeping fixed $a$ and $Q$. Note that we have set $\xi\to 0$ for maximum energy extraction.}} 
\end{center}
\end{figure*}
\begin{figure*}
    \centering
    \includegraphics[scale=0.5]{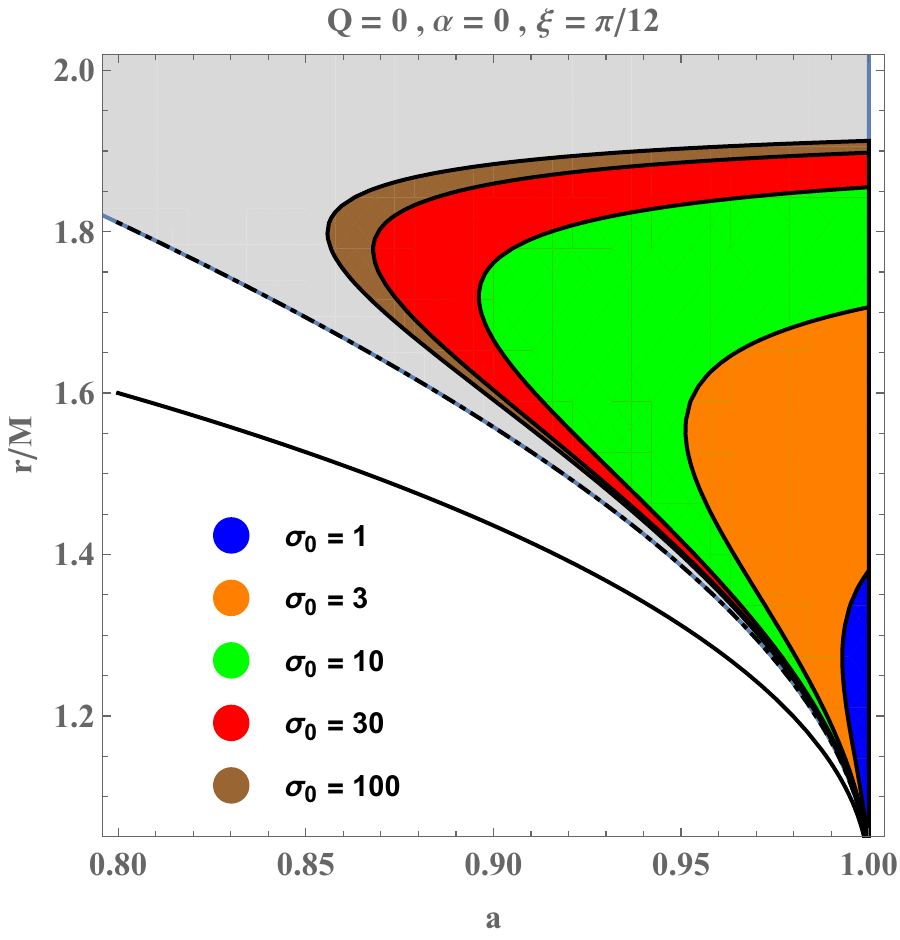}
    \includegraphics[scale=0.5]{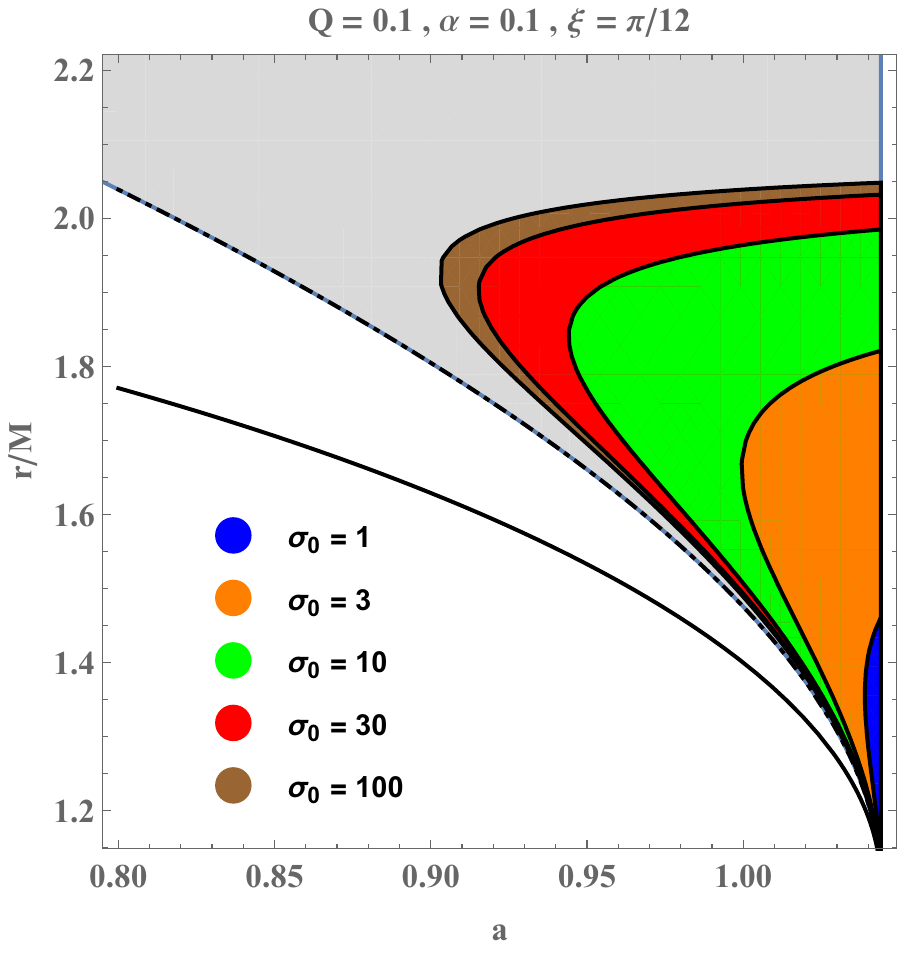}
    \includegraphics[scale=0.5]{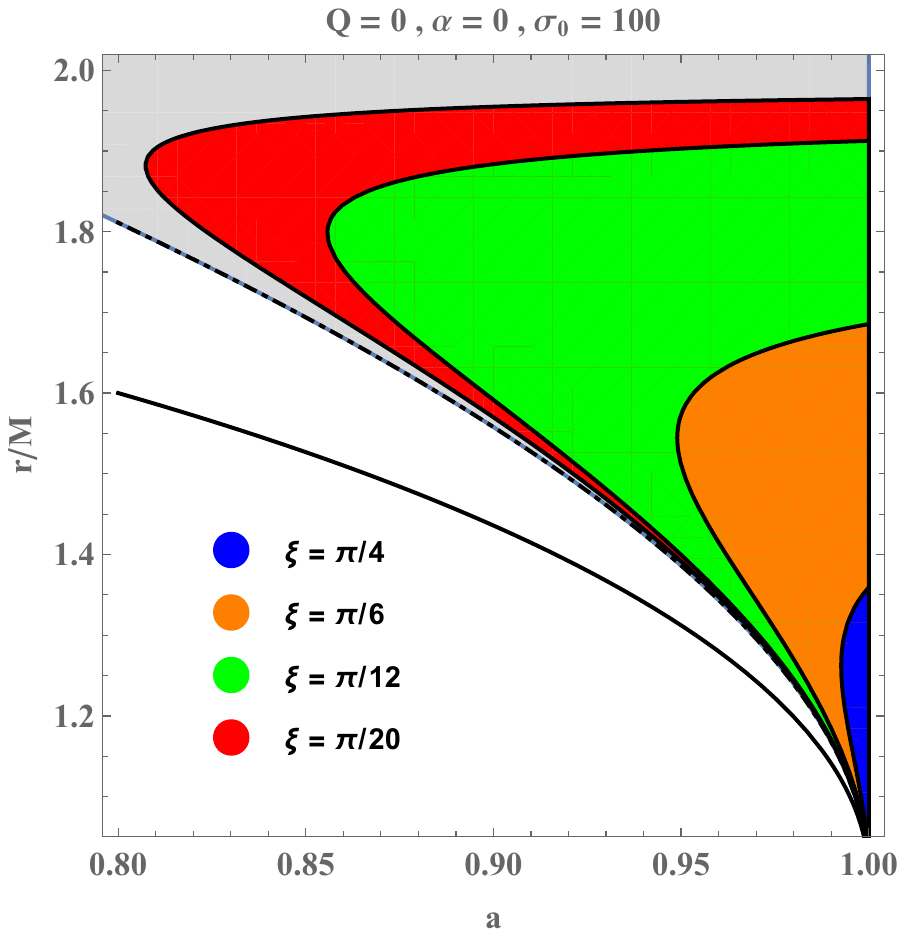}
    \includegraphics[scale=0.5]{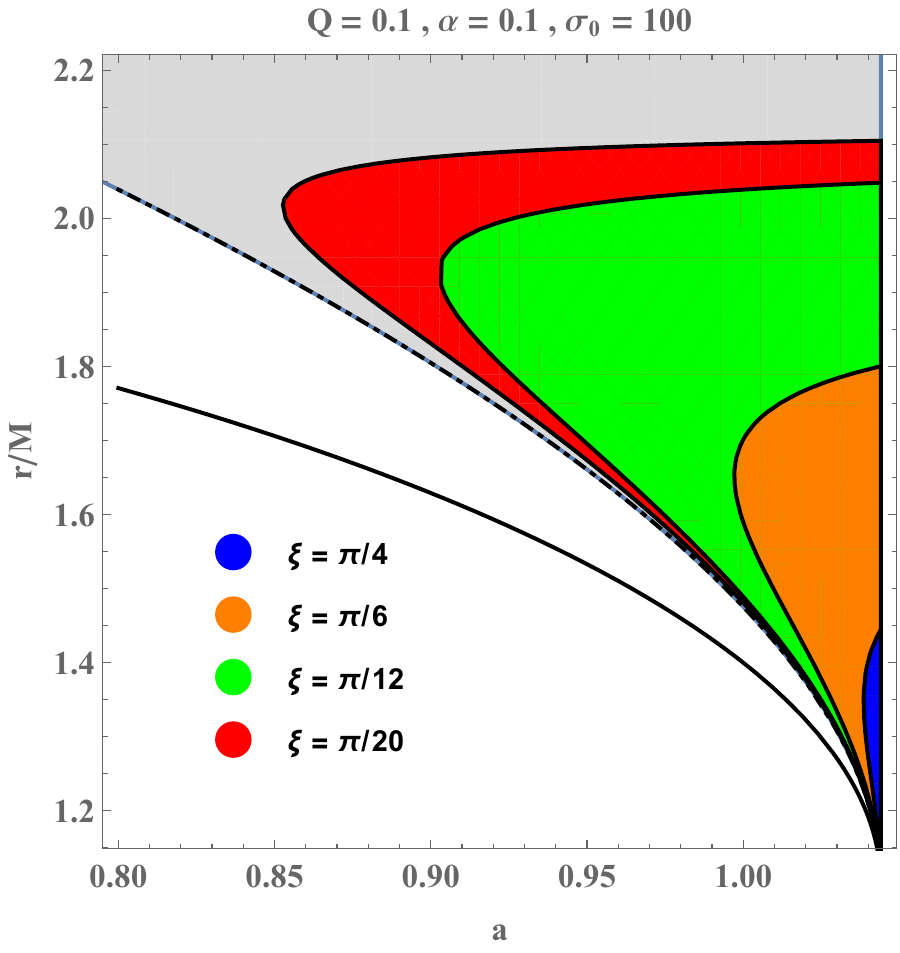}
    \caption{\label{fig:phase-space}The phase-space regions \{$a$, $r/M$\} are plotted for the accelerated plasma energy $\epsilon^{\infty}_{+}>0$ (gray area) and the decelerated plasma energy $\epsilon^{\infty}_{-}<0$ (blue to red areas). Top row, left/right panels: \{$a$, $r/M$\} is plotted for various combinations of the plasma magnetization $\sigma_0$ in the case of Kerr/Kerr-Newman-MOG black hole cases. Bottom row, left/right panels: \{$a$, $r/M$\} is plotted for various combinations of the orientation angle $\xi$ in the case of Kerr/Kerr-Newman-MOG black hole cases. }
\end{figure*}

Black holes are most fascinating objects as that of their rich astrophysical energetic phenomena, i.e.,  the outflows coming from active galactic nuclei with the energy range of order $E\approx 10^{42}$-$10^{47}\:\rm{erg/s}$ observed with the help of x rays, $\gamma$ rays, and very long baseline interferometry observations~\cite{Fender04mnrs,Auchettl17ApJ,IceCube17b}. These outflows in the form of winds and jets are related to the charged particle motion moving on the accretion disk around the black hole. {To explore this energetic phenomenon we analyse the magnetic reconnection process by applying the recently proposed Comisso-Asenjo mechanism~\cite{Comisso21}, which strongly depends on the frame dragging effect twisting the magnetic field lines around a rapidly spinning black hole. We further consider that the magnetic reconnection takes place in the ergoregion of a rapidly spinning Kerr-Newman MOG black hole so that its rotational energy is driven out efficiently by this scenario.} For that, we first consider the zero angular momentum observer (ZAMO) frame and then we examine the plasma energy density. For the above mentioned ZAMO frame the line element is written as follows:
  \begin{eqnarray}
      ds^2=-d\hat{t}^2+\sum_{i=1}^3(d\hat{x^i})^2=g_{\mu \nu}dx^{\mu}dx^{\nu}\, , 
  \end{eqnarray}
where we define $d\hat{t}$ and $d\hat{x^i}$ as follows: 
\begin{eqnarray}
d\hat{t}=\alpha dt \mbox{~~and~~} d\hat{x^i}=\sqrt{g_{ii}}dx^i-\alpha \beta^{\phi}dt\, ,
\end{eqnarray}
with $\alpha$ and $\beta^i=(0,0,\beta^{\phi})$ that, respectively, refer to the laps function and the shift vector and are written as 
\begin{eqnarray}
\alpha=\sqrt{-g_{tt}+\frac{g_{\phi t}^2}{g_{\phi \phi}}}  \mbox{~~and~~}\beta^{\phi}=\frac{\sqrt{g_{\phi\phi}} \omega^{\phi}}{\alpha}\, .
\end{eqnarray}
Note here that $\omega^{\phi}=-{g_{\phi t }}/{g_{\phi \phi}}$ is referred to as the frame dragging velocity of the zero angular momentum particle. For the ZAMO frame the vector
$\psi$ has the following covariant and contravariant components: 
\begin{eqnarray}\label{Boyer}
\hat{\psi_0}=\frac{\psi_0}{\alpha}+\sum_{i=1}^3 \frac{\beta^i}{g_{ii}}\psi_i \mbox{~~and~~} \hat{\psi_i}=\frac{\psi_i}{\sqrt{g_{ii}}}\, ,\\
\hat{\psi^0}=\alpha \psi^0 \mbox{~~and~~} \hat{\psi^i}=\sqrt{g_{ii}}\psi^i-\alpha \beta^i \psi^0\, .
\end{eqnarray}
It is well known that the ergosphere, where $\epsilon = -p.\partial/\partial t$, becomes negative, thus resulting in having negative energy for a timelike particle. This plays a key role in energy extraction by the magnetic reconnection process that can be evaluated in the region between static surface $r_{st}$ and the horizon $r_{+}$. In the case of the Kerr-Newman-MOG black hole the ergosphere is illustrated in Fig.~\ref{fig:ergo}. For this process, the key property is to accelerate the plasma with high energy. As a consequence of the magnetic reconnection process, the accelerated plasma comes out with arbitrarily high energy. Otherwise, it is  decelerated and it then attains negative energy so as to get absorbed by the black hole. To be more accurate we define the energy-momentum tensor for the one-fluid approximation of the plasma, and it is given by  
\begin{eqnarray}
    T^{\mu \nu}=pg^{\mu \nu}+{\mathit{w}}U^{\mu}U^{\nu}+F^{\mu}_{\delta}F^{\nu \delta}-\frac{1}{4}g^{\mu \nu}F^{\rho \delta}F_{\rho \delta}\, .
\end{eqnarray}
Note that in the above equation $p$ and $\mathit{w}$, respectively, refer to the proper plasma pressure and enthalpy density, while $U^{\mu}$ and $F^{\mu \nu}$ refer to four-velocity and the tensor of the electromagnetic field. Here we note that the enthalpy density is given by $\mathit{w}=e_{int}+p$, where the thermal energy density  is defined by \cite{Koide08ApJ} 
\begin{equation}
    e_{int}=\dfrac{p}{\Gamma-1}+\rho c^2\, ,
\end{equation}
with $\Gamma$ and $\rho$ which, respectively, refer to the adiabatic index and the proper mass density. By imposing this condition we further define the relativistic hot plasma with the equation of state.  
The energy density at infinity can be given by the following relation  \begin{eqnarray}
e^{\infty}=-\alpha g_{\mu 0}T^{\mu 0}\, .
\end{eqnarray}
Taking this into account, one can write the energy density at infinity as follows:    
\begin{equation}
    e^{\infty}=\alpha \hat{e}+\alpha \beta^{\phi}\hat{P}^{\phi}\, ,
\end{equation}
where $\hat{e}$ and $\hat{P}^{\phi}$,  respectively, define the total energy density and azimuthal component of the momentum density, and they are given by 
\begin{eqnarray}
    \hat{e}={\mathit{w}}\hat{\gamma}^2-p+\frac{\hat{B}^2+\hat{E}^2}{2}\, ,\\
   \hat{P}^{\phi}={\mathit{w}}\hat{\gamma}^2\hat{v}^{\phi}+(\hat{B} \times \hat{E})^{\phi}\, ,
\end{eqnarray}
where $\hat{v}^{\phi}$ represents the azimuthal component of the plasma velocity at the ZAMO. The Lorentz factor $\hat{\gamma}$ and electric $\hat{E}^i$ and magnetic $\hat{B}^i$ field components that appear in the above equation are defined by 
\begin{eqnarray}
\hat{\gamma}=\hat{U}^0=\sqrt{1-\sum_{i=1}^3(d\hat{v}^i)^2}\, ,\\ 
\hat{B}^i=\epsilon^{ijk}\hat{F}_{jk}/2 \mbox{~~and~~} \hat{E}^i=\eta^{ij}\hat{F}_{j0}=\hat{F}_{i0}\, .
\end{eqnarray}
It is worth noting here that the energy density at infinity $e^{\infty}$ consists of two parts, i.e., the hydrodynamic and electromagnetic parts with  $e^{\infty}=e_{hyd}^{\infty}+e_{em}^{\infty}$ which can be written separately as follows:
\begin{eqnarray}
 e_{hyd}^{\infty}=\alpha \hat{e}_{hyd}+\alpha \beta^{\phi}{\mathit{w}}\hat{\gamma}^2\hat{v}^{\phi}\, ,   \\
 e_{em}^{\infty}=\alpha \hat{e}_{em}+\alpha \beta^{\phi}(\hat{B}\times \hat{E})_{\phi}\, , 
\end{eqnarray}
where  $\hat{e}_{hyd}={\mathit{w}}\hat{\gamma}^2-p$ and $\hat{e}_{em}=(\hat{B}^2+\hat{E}^2)/2$, respectively, refer to the energy densities of the hydrodynamic and electromagnetic fields at the ZAMO.  We then need to evaluate the energy density at infinity. For that, we assume that the contribution of the electromagnetic field can be expelled out since its effect is very small in contrast to the hydrodynamic energy density at infinity. However, we note that most of the magnetic field energy can be transferred to the plasma kinetic energy in the magnetic reconnection process. Taking all together we shall further assume that we apply for incompressible and adiabatic plasma for the approximation. In doing so, the energy density at the infinity is then defined by the following form~\cite{Comisso21}:
\begin{equation}
    e^{\infty}=e^{\infty}_{hyd}=\alpha {\mathit{w}}\hat{\gamma}(1+\beta^{\phi}\hat{v}^{\phi})-\frac{\alpha p}{\hat{\gamma}}\, .
\end{equation}
Next, one needs to make the magnetic reconnection process the localized one so as to determine it at a small scale. For that, the local rest frame $x'^{\mu}=(x'^{0},x'^{1},x'^{2},x'^{3})$ comes into play because of the bulk plasma that orbits at the equatorial plane around the black hole with the Keplerian frequency/angular velocity $\Omega_K$, which can be defined by 
\begin{equation}\label{angular1}
    \Omega_K=\dfrac{d \phi}{d t}=\dfrac{-g_{t \phi,r}+\sqrt{g^2_{t \phi,r}-g_{tt,r}g_{\phi \phi,r}}}{g_{\phi \phi,r}}\, .
\end{equation}
The Keplerian frequency for the Kerr-Newman-MOG black hole then reads as follows:  
\begin{eqnarray}\label{Eq:KF}
    \Omega_K&=&\frac{a \Big(Q^2-(\alpha +1) (r-\alpha )\Big)}{r^4+a^2 \Big(Q^2-(\alpha +1) (r-\alpha )\Big)}\nonumber\\&+&\frac{r^2 \sqrt{(\alpha +1) (r-\alpha )-Q^2}}{r^4+a^2 \Big(Q^2-(\alpha +1) (r-\alpha )\Big)}\, .
\end{eqnarray}
It is worth noting that the direction of $x'^{\mu}$ is chosen so that $x'^{1}$ and $x'^{3}$ must be parallel to the radial $x^1=r$ and the azimuthal $x^3=\phi$ directions, respectively. Following Eq.~(\ref{Boyer}) we further consider the corotating Keplerian frequency at the ZAMO, and it is given by 
\begin{eqnarray}
 \hat{v}_K&=&\frac{d\hat{x}^{\phi}}{d\hat{x}^{t}}%=\frac{d\hat{x}^{\phi}/d\lambda}{d\hat{x}^{t}/d\lambda}
 =\frac{\sqrt{g_{\phi \phi}}dx^{\phi}/d\lambda-\alpha\beta^{\phi}dx^t/d\lambda}{\alpha dx^t/d\lambda} \nonumber \\
 &=&\frac{\sqrt{g_{\phi \phi}}}{\alpha}\Omega_K-\beta^{\phi}\, . 
 \end{eqnarray}
One can then obtain the forms 

of $\hat{v}_K$ and the Lorentz factor $\hat{\gamma}_K=1/\sqrt{1-\hat{v}_K^2}$ by imposing the Keplerian frequency $\Omega_K$ given by Eq.~(\ref{Eq:KF}).  
The rotation energy of a black hole that can be extracted by the magnetic reconnection process extremely depends upon the plasma dynamics and electromagnetic
filed properties. As was mentioned, we assume a one-fluid plasma that obeys adiabatic and incompressible plasma approximation so that the hydrodynamic energy per enthalpy at the infinity reads as follows \cite{Comisso21}:
\begin{eqnarray}
 \epsilon^{\infty}_{\pm}&=&\alpha \hat{\gamma}_K \Bigg[(1+\beta^{\phi}\hat{v}_K)\sqrt{1+\sigma_0}\pm \cos{\xi}(\beta^{\phi}+\hat{v}_K)\sqrt{\sigma_0}\nonumber\\
 &-&\frac{\sqrt{1+\sigma_0}\mp \cos{\xi}\hat{v}_K\sqrt{\sigma_0}}{4\hat{\gamma}^2(1+\sigma_0-\cos^2{\xi}\hat{v}_K^2\sigma_0)}\Bigg]\, , 
\end{eqnarray}
with $\sigma_0=B_0^2/{\mathit{w}}$ and $\xi$, which, respectively, refer to the plasma magnetization and the orientation angle between the magnetic field and the outflow plasma directions at the equatorial plane. To extract the black hole rotational energy on the basis of the magnetic reconnection process the hydrodynamic energy would be positive when the plasma is accelerated. In contrast, the energy is negative when it is decelerated near the black hole horizon, similar to what is observed for the Penrose process. Thus, the energy that can be extracted by the magnetic reconnection process should be positive and always much greater than thermal energies and the rest mass of the plasma as well. We shall further assume that the plasma is a relativistic hot plasma that satisfies the condition for the equation of state, i.e., $\mathit{w} = 4p$ \cite{Comisso21}. With this in view,  the accelerated and decelerated energies of the plasma which can be measured at infinity read as follows: 
\begin{eqnarray}
 \epsilon_{-}^{\infty}<0 \mbox{~~and~~} \Delta \epsilon_{+}^{\infty}>0\, ,
 \end{eqnarray}
 where $\Delta \epsilon_{+}^{\infty}$ is given by 
 \begin{eqnarray}
 \Delta \epsilon_{+}^{\infty}=\epsilon_{+}-\left(1-\frac{\Gamma}{\Gamma-1}\frac{p}{\mathit{w}}\right)>0\, .
\end{eqnarray}
Taking the polytropic index as $\Gamma$=4/3 allows one to have the form as $ \Delta \epsilon_{+}^{\infty}=\epsilon_{+}^{\infty}>0$ for the relativistic hot plasma.
\begin{figure*}
    \centering
    \includegraphics[scale=0.45]{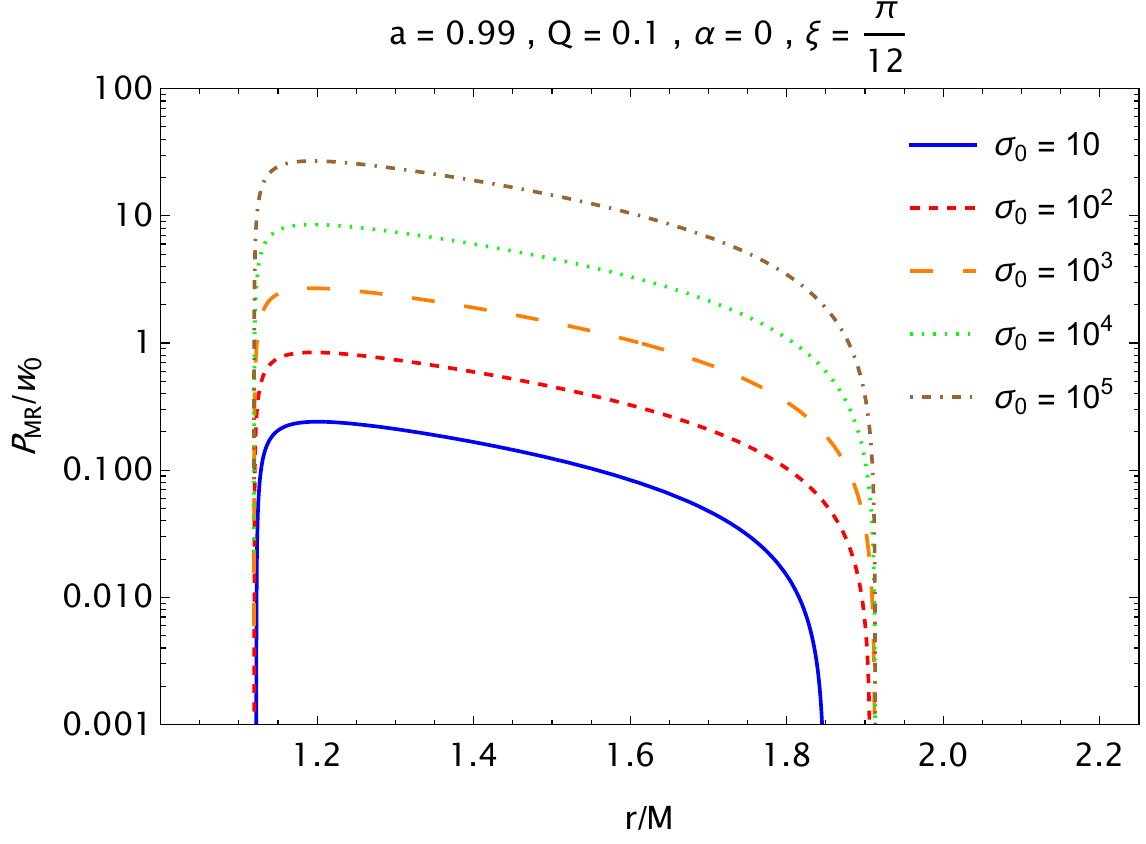}
    \includegraphics[scale=0.45]{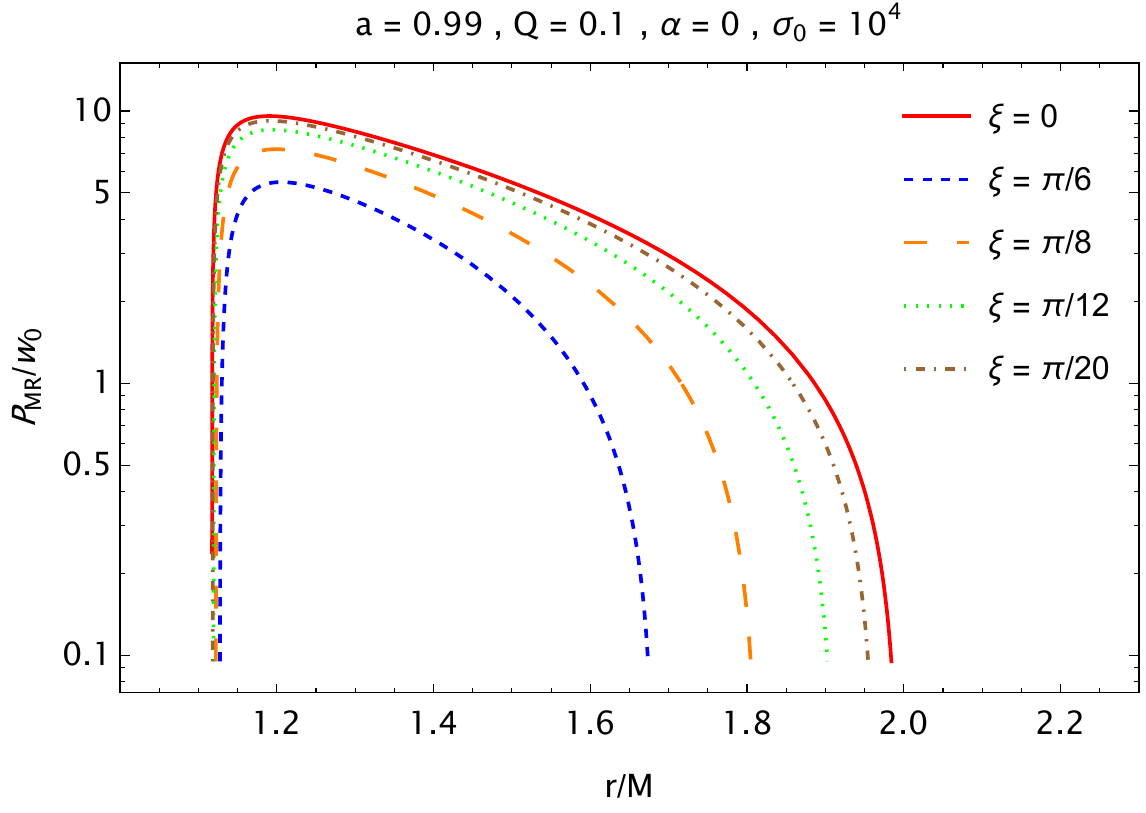}
    \includegraphics[scale=0.45]{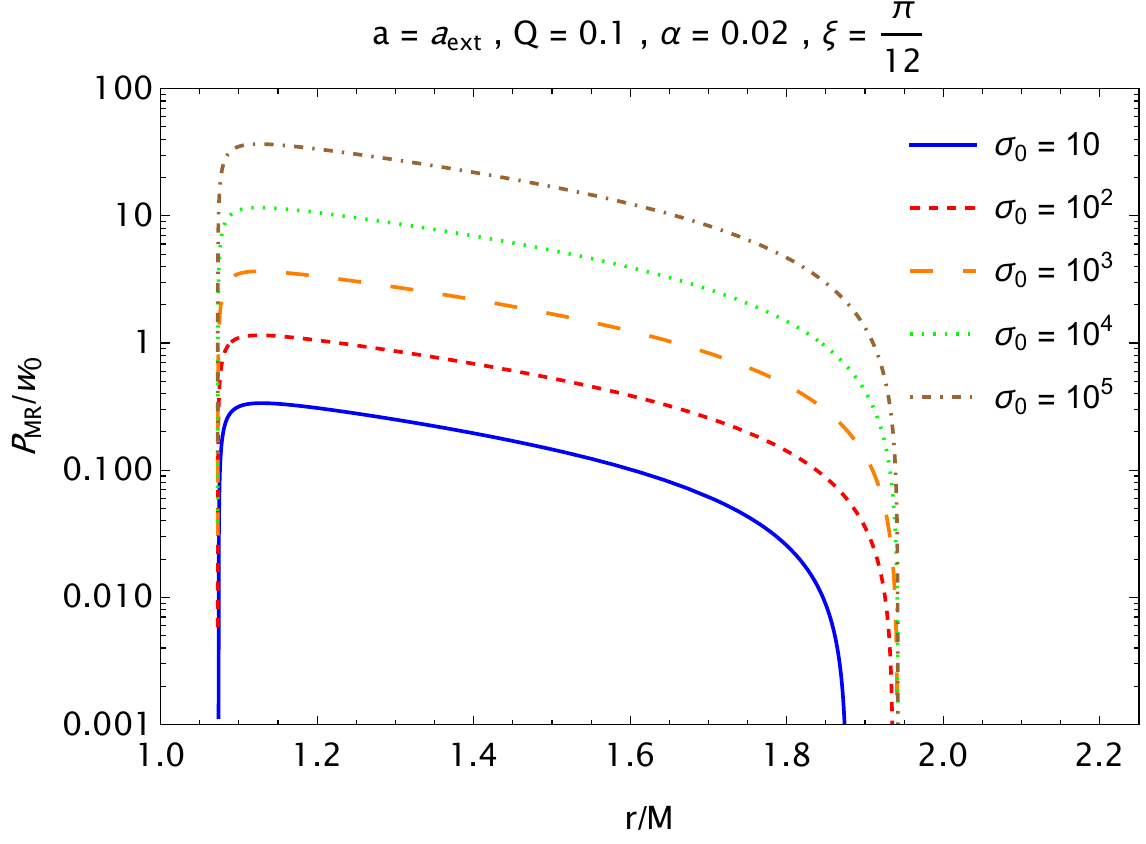}
    \includegraphics[scale=0.45]{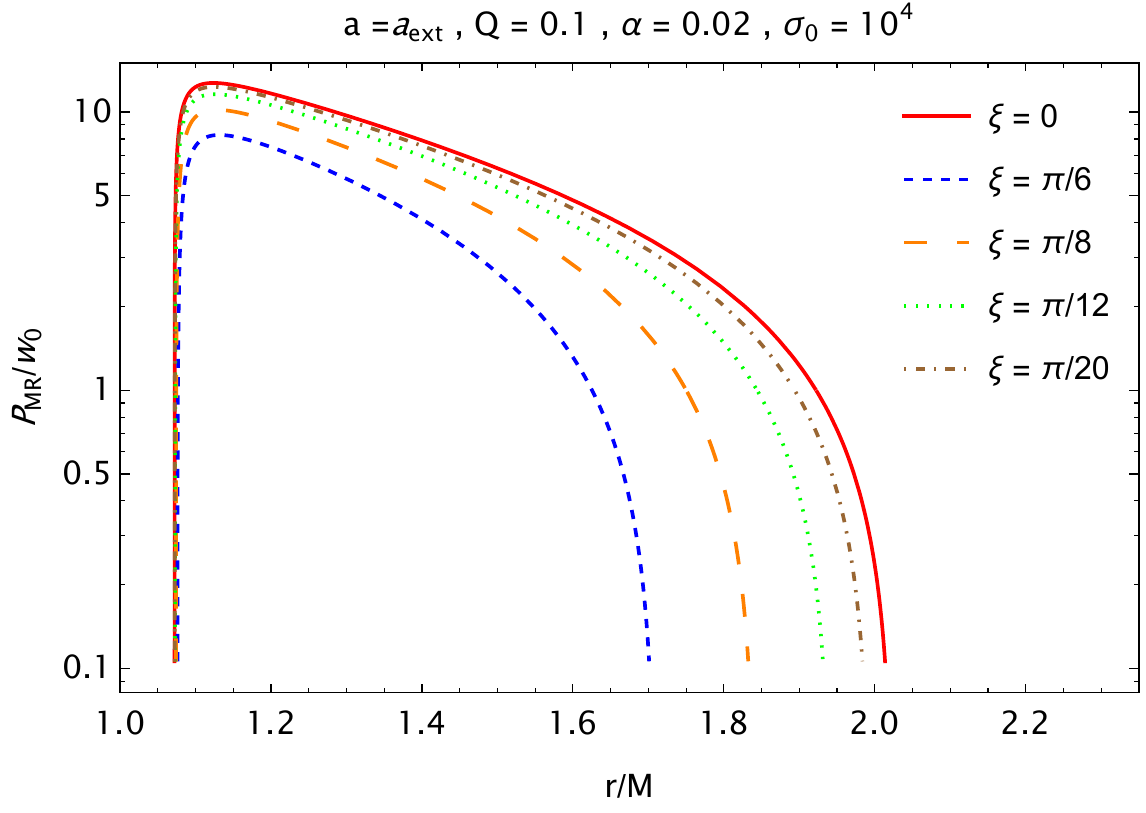}
     \includegraphics[scale=0.45]{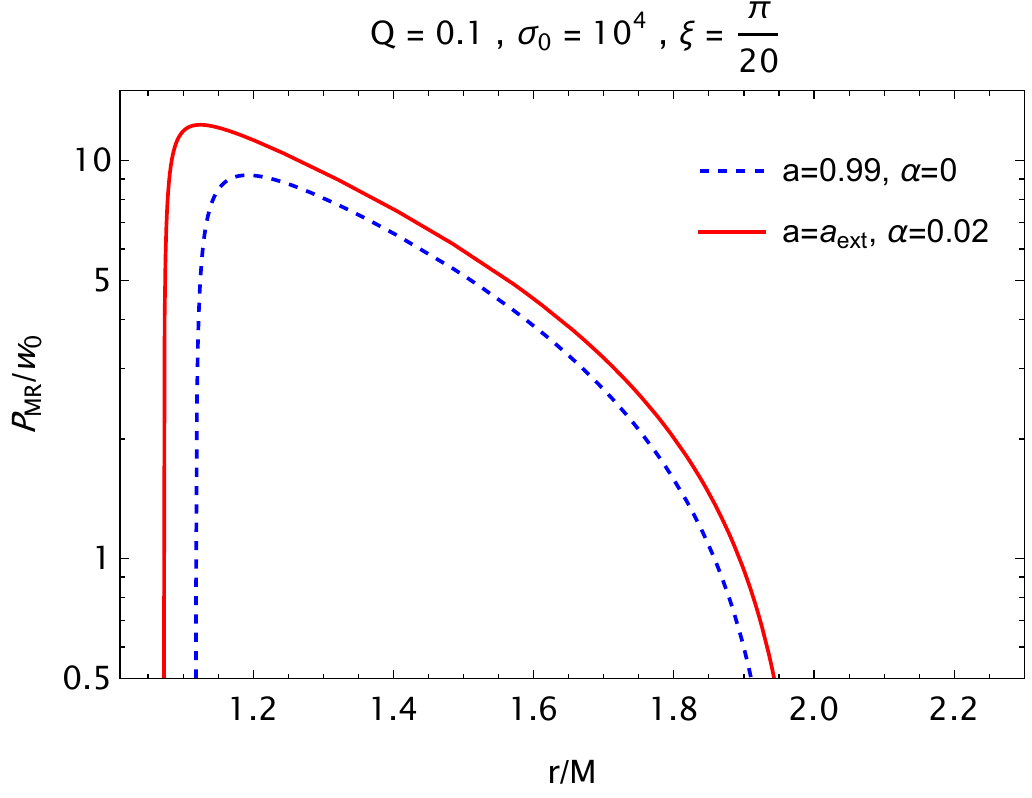}
            \caption{\label{fig:power}The power $P_{MR}/\mathit{w}_0$ of the magnetic reconnection is plotted for a rapidly spinning Kerr-Newman-MOG black hole. Top row, left: $P_{MR}/\mathit{w}_0$ is plotted for various combinations of the plasma magnetization $\sigma_0$ while keeping fixed $a$, $Q$, and $\xi$. Top row, right: $P_{MR}/\mathit{w}_0$ is plotted for various combinations of the orientation angle $\xi$ while keeping fixed $a$, $Q$, and $\sigma_0$. Middle row similar to what is observed in the top row, $P_{MR}/\mathit{w}_0$ is plotted as a consequence of the presence of MOG parameter $\alpha$. Bottom row: $P_{MR}/\mathit{w}_0$ is plotted for Kerr and Kerr-Newman-MOG black hole cases while keeping fixed $Q=0.1$, $\sigma_0=10^{4}$, and $\xi=\pi/20$. %\cred
            {Note that an extremal value of the spin parameter refers to $a_{ext}=(1+\alpha-Q^2)^{1/2}$; see Fig.~\ref{fig:aext}.}   }
\end{figure*}

We now analyze the accelerated $\epsilon_{+}^{\infty}$ and decelerated $\epsilon_{-}^{\infty}$ energies of the plasma  to explore the remarkable aspects of energy extraction through the magnetic reconnection process for the Kerr-Newman-MOG black hole. However, the analytic forms of $\epsilon_{+}^{\infty}$ and $\epsilon_{-}^{\infty}$ energies turn out to be very long and complicated expressions for explicit display. %\cred{We therefore further resort to numerical evaluation of these plasma energies.} 
 In Fig.~\ref{Fig:energy}, we therefore show the plasma energies $\epsilon_{+}^{\infty}$ and $\epsilon_{-}^{\infty}$ as a function of the plasma magnetization parameter for various possible cases. As seen in Fig.~\ref{Fig:energy}, the left panel shows the impact of the black hole spin parameter $a$ on the plasma magnetization profile of the accelerated and decelerated energies for $Q=0$ and $\alpha=0$, while, similarly, the right panel shows the impact of the MOG parameter $\alpha$ for fixed $a=0.99$ and $Q=0.1$. As shown in Fig.~\ref{Fig:energy}, the accelerated and decelerated energies per enthalpy increase as the MOG parameter is taken into account for spin parameter $a=0.99$, thus resulting in compensating the impact of spin parameter $a$ on these energies, i.e, $\epsilon_{+}^{\infty}$ and $\epsilon_{-}^{\infty}$. This happens because the spin parameter turns out to be $a>1$ as a consequence of the presence of $\alpha$. Unlike the Kerr black hole, the maximum energy is not restricted by $a=1$ for the Kerr-Newman MOG black hole case, thus leading to arbitrarily high energy that can be extracted by the magnetic reconnection process as a consequence of an increase in the value of MOG parameter $\alpha$. We also further explore another important parameter region ($a, r/M$), which we refer to as the phase-space region for the energy extraction through the magnetic reconnection process. In the phase-space region, the conditions $\epsilon_{+}^{\infty}>0$ and $\epsilon_{-}^{\infty}<0$ are always satisfied so that the energy can be extracted from the black hole, i.e., $\eta=\epsilon_{+}^{\infty}/(\epsilon_{+}^{\infty}+\epsilon_{-}^{\infty})>1$. Figure~\ref{fig:phase-space} reflects the role of the magnetization parameter $\sigma_0$ (top row, left and right panels) and the orientation angle $\xi$ (bottom row, left and right panels) on the regions of the phase space ($a, r/M$). From the top row panels of Fig.~\ref{fig:phase-space}, it is clearly seen that the phase-space region, in which the energy can be extracted via the magnetic reconnection, shifts upward to a larger location $r$ and to smaller values of the spin parameter $a$ as a consequence of an increase in the value of magnetization parameter $\sigma_0$ of the plasma for fixed orientation angle $\eta=\pi/12$. It turns out that the large values of $\sigma_0$ give rise to an expanded phase-space region that, in turn, leads to arbitrarily high energy extracted by the magnetic reconnection process. Similarly, the phase-space region gets also influenced strongly by the orientation angle $\xi$ of the reconnecting magnetic field, as seen in the bottom panels of Fig.~\ref{fig:phase-space}. That is, the phase-space region for energy extraction can extend toward up to larger $r$ and to lower $a$ by decreasing the orientation angle $\eta$ for fixed magnetization parameter $\sigma_0=100$. This happens because a small orientation angle would lead to making the azimuthal component of the outflow plasma velocity dominate over its rest components, thus resulting in giving the main contribution to the energy extraction process. The most important key point to be noted here is that the combined effect of black hole charge and MOG parameter can extend the phase space for the energy extraction condition, i.e., $\delta \epsilon^{\infty}_{+}>0$ (gray area), as seen in both top and bottom right panels of Fig.~\ref{fig:phase-space}.       

Next, we investigate the power and energy efficiency via the magnetic reconnection process for the Kerr-Newman-MOG black hole.

\begin{figure*}
    \centering
    \includegraphics[scale=0.49]{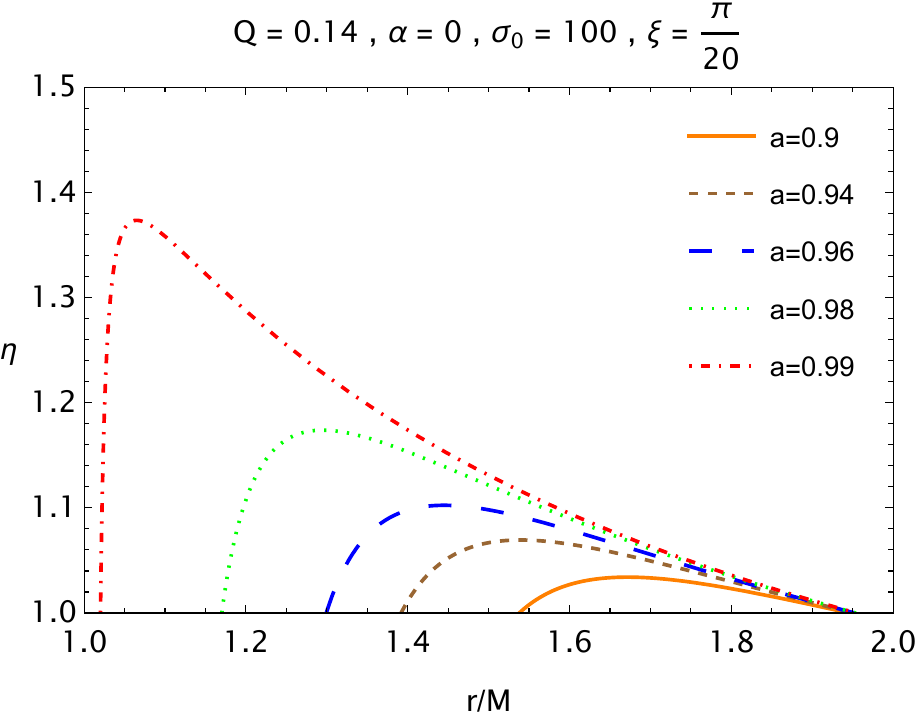}
     \includegraphics[scale=0.4]{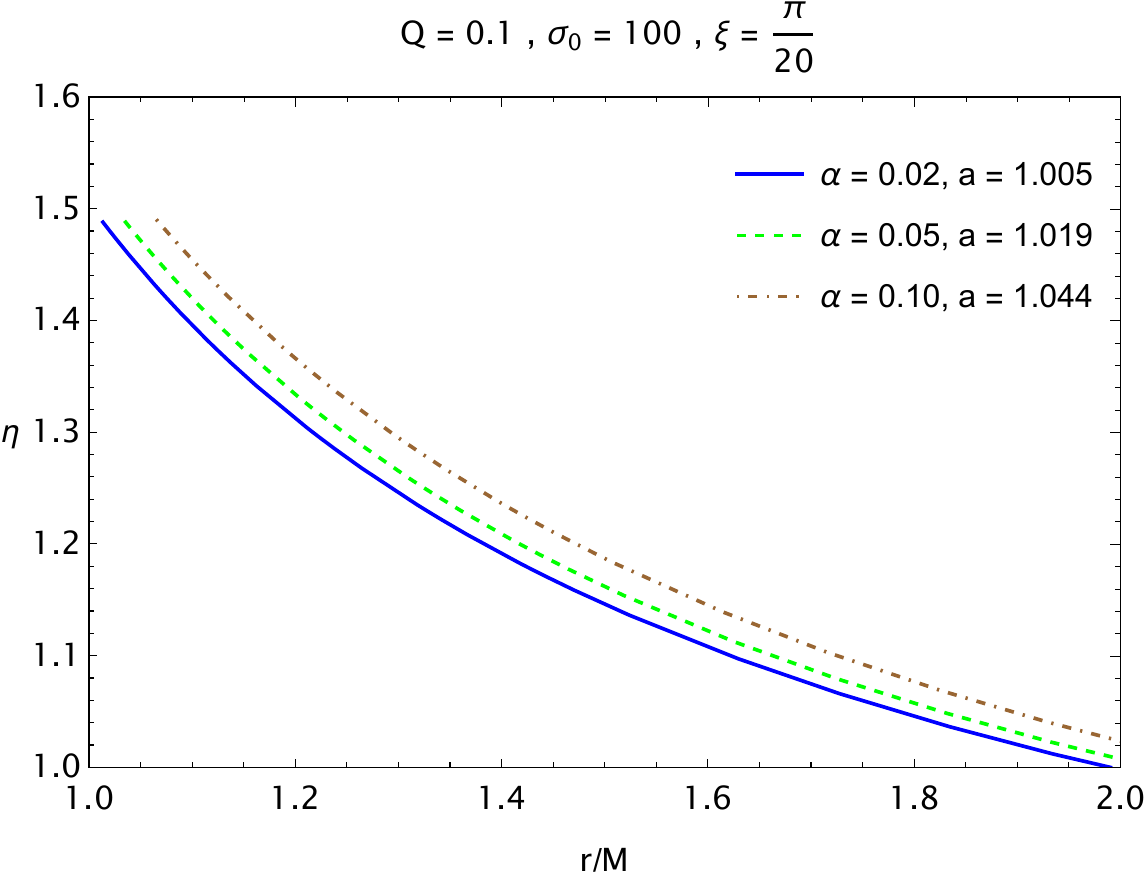}
    \caption{\label{fig:efficiency}The magnetic reconnection efficiency against the location $r/M$ for various combinations of spin parameter (left) and MOG parameter $\alpha$ (right) in the case of fixed black hole charge $Q$. Here we have set $\sigma_0=100$ for the plasma magnetization $\xi=\pi/20$ for the orientation angle the reconnecting magnetic can have.  }
   \end{figure*}
\begin{figure*}
    \centering
    \includegraphics[scale=0.45]{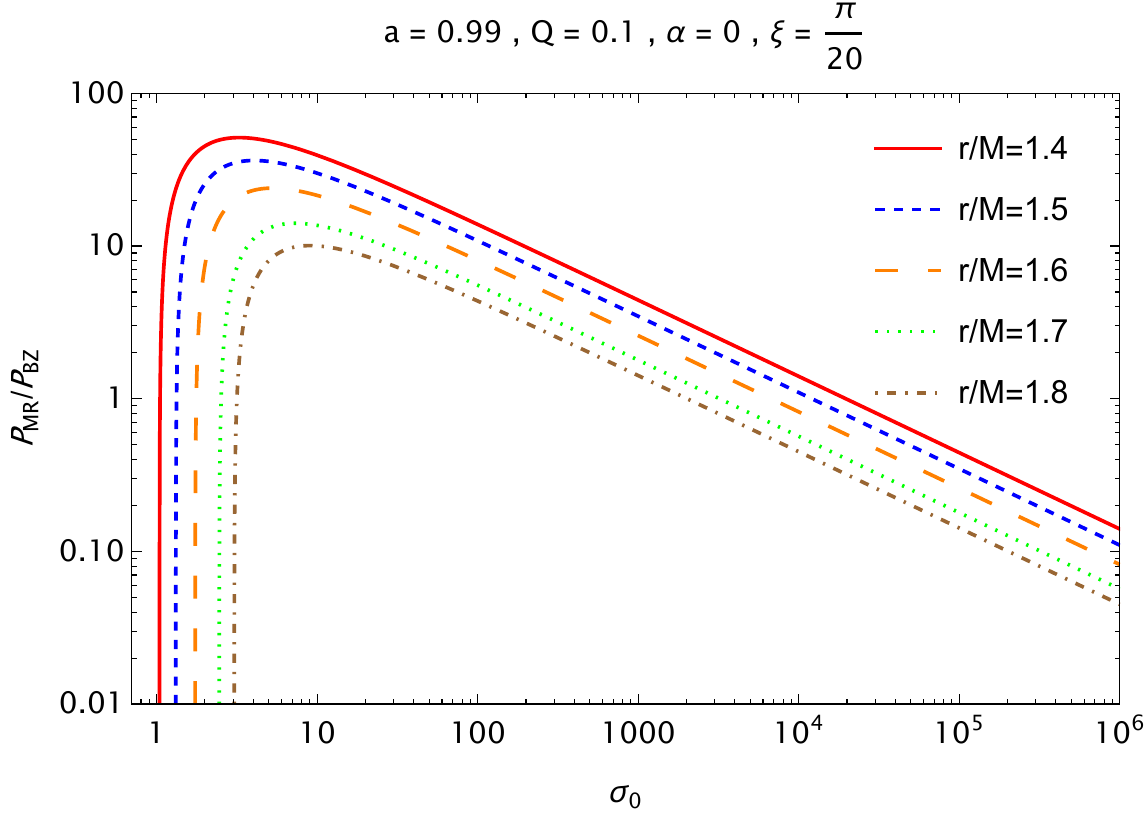}
    \includegraphics[scale=0.45]{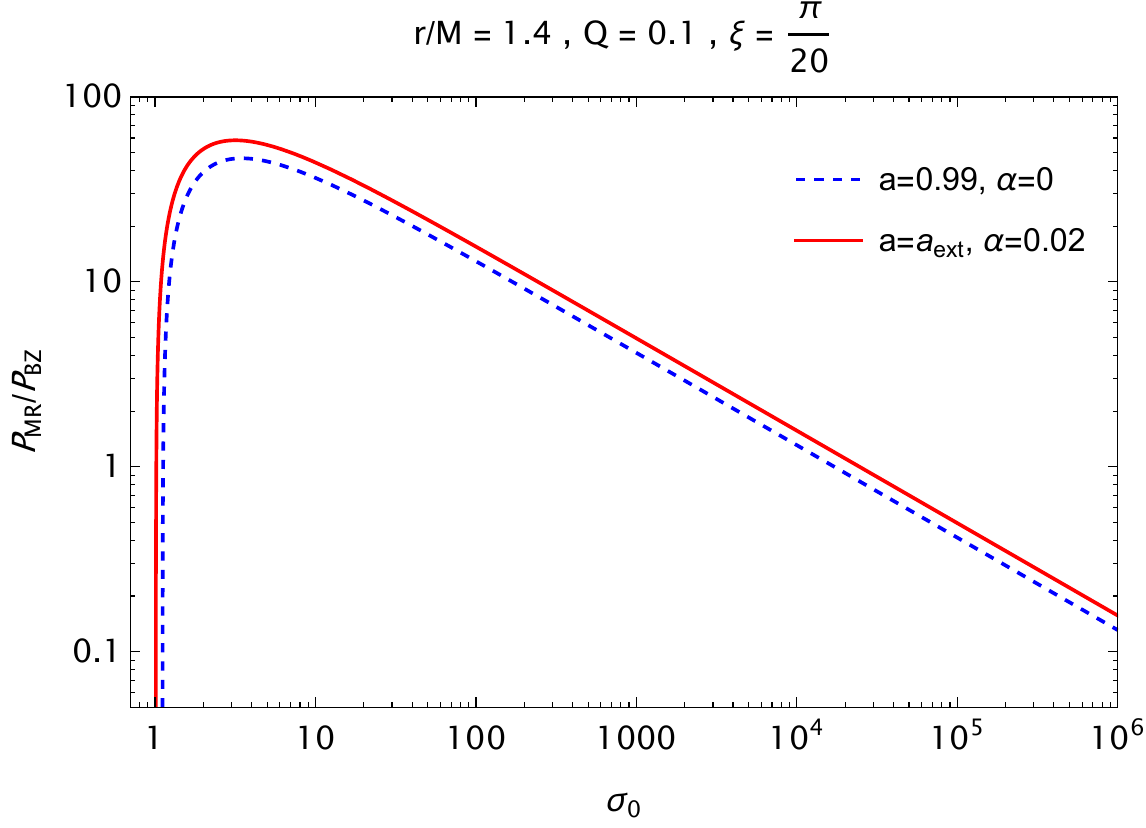}
    \caption{\label{fig:ratio}The ratio of power $P_{MR}/P_{BZ}$ against the plasma magnetization $\sigma_0$. Left: $P_{MR}/P_{BZ}$ is plotted for various combinations of location $r/M$ for fixed values of black hole parameters and the orientation angle. Right: $P_{MR}/P_{BZ}$ is plotted for two values of MOG parameter $\alpha$ and corresponding values of spin parameter. }
\end{figure*}

\section{\label{Sec:power-mm}POWER AND MAGNETIC RECONNECTION EFFICIENCY}

In this section, following the Comisso-Asenjo mechanism \cite{Comisso21} originally suggested for the Kerr black hole, we consider the power and the energy efficiency via the magnetic reconnection process and resort to the numerical evaluation of these quantities for the Kerr-Newman-MOG black hole. It is worth noting that energy efficiency and power are strongly related to the negative energy of decelerated plasma which is absorbed by a black hole in the unit time. We first turn to the power of the energy extraction in which it is envisaged that the power that is extracted from a black hole due to the escaping plasma can be defined by \cite{Comisso21} 
\begin{eqnarray}
    P_{MR}=-\epsilon_{-}^{\infty}{\mathit{w}}_0 A_{in} U_{in}\, ,
\end{eqnarray}
where $U_{in}$ defines the regime, i.e., $U_{in}={\cal O}(10^{-1})$ refers to  the collisionless regime, while $U_{in}={\cal O}(10^{-2})$ to the collisional regime. Also, $A_{in}$ given in the above expression defines the cross sectional area for the inflowing and is evaluated as $A_{in}=(r^2_E-r^2_{ph})$ for rotating black holes. Note that $r_E$ is given by Eq.~(\ref{Eq:Ergo}), while we resort to the numerical estimation of $r_{ph}$ which can tend to $r_{H}$ at the extremal black hole case.      

 In Fig.~\ref{fig:power}, we show the radial profile of the power, $P_e/{\mathit{w}}_0$, that can be extracted from the black hole by the outflowing plasma generated by the magnetic reconnection. As can be seen from Fig.~\ref{fig:power}, the panel in the top row reflects the role of the magnetization parameter $\sigma_0$ for the radial profile of the power while keeping black hole spin $a$, charge $Q$, and MOG parameter $\alpha$ fixed and the orientation angle $\xi=\pi/12$, while the right panel reflects the role of $\xi$ in the case in which the parameters $a$, $Q$,  $\alpha$, and $\sigma_0=10^{4}$ are fixed. %\cred
 {Similarly, the panels in the middle row of Fig.~\ref{fig:power} show similar behavior of the radial profile of the power as a consequence of the presence of the MOG parameter $\alpha$ that would lead to higher spin $a>1$ relative to the Kerr black hole case; see Figs.~\ref{fig:Region_Plot_btw_alpha_n_a} and \ref{fig:aext}}. As seen in the top row of Fig.~\ref{fig:power}, the height of the power extracted from the black hole increases and its shape shifts toward up to its higher values as the magnetization parameter $\sigma_0$ and the orientation angle $\eta$ increase. We also show that the inclusion of the MOG parameter $\alpha$ makes the power more effective, thus resulting in shifting its height toward up to larger values of the power, as seen in the middle row of Fig.~\ref{fig:power}. We must ensure that the inclusion of the MOG parameter $\alpha$ leads to arbitrarily high power extracted by the magnetic reconnection. To that, we show the combined effect of black hole charge and MOG parameter in the bottom panel of Fig.~\ref{fig:power}. One can then infer that the combination of black hole charge and the MOG parameter $\alpha$ leads to arbitrarily high power through the magnetic reconnection.       
 
Let us then turn to another key aspect of energy extraction via magnetic reconnection.  To understand how magnetic reconnection is an effective model, one needs to analyze the energy release that can be extracted by the model considered here. Therefore, we need to evaluate the total amount of energy released by the Kerr-Newman-MOG black hole. We first would like to emphasize that the magnetic field energy during the magnetic reconnection, after being distributed, consists of two parts of energies, i.e., the decelerated plasma and the accelerated plasma energies The former one having negative energy is absorbed by the black hole, while the latter one with arbitrarily high positive energy comes out toward to larger $r$ distances from the black hole. With this in view,  the energy efficiency of the plasma through the magnetic reconnection can generally be 
defined by 
\begin{eqnarray}
    \eta=\frac{\epsilon_{+}^{\infty}}{\epsilon_{+}^{\infty}+\epsilon_{-}^{\infty}}\, ,
\end{eqnarray}
where $\epsilon_{+}^{\infty}$ and $\epsilon_{-}^{\infty}$, respectively, define the accelerated and decelerated plasma 
energies, as mentioned above. The key point to be noted here is that for the plasma energy to be released from a black hole the condition  $\eta=\epsilon_{+}^{\infty}/(\epsilon_{+}^{\infty}+\epsilon_{-}^{\infty})>1$ must always be satisfied for the energy efficiency. We further then analyze the energy efficiency released by the magnetic reconnection.   

In Fig.~\ref{fig:efficiency}, we show the radial profile of the energy efficiency for the outflowing plasma generated by the magnetic reconnection for various combinations of black hole spin parameter $a$ and the MOG parameter $\alpha$. It is easily notable from the left panel of Fig.~\ref{fig:efficiency} that the energy efficiency grows as a function of $r/M$ in the close vicinity of the black hole {(i.e., in the ergosphere of the black
hole)} as spin parameter $a$ increases for fixed black hole charge $Q=0.14$ and the orientation angle $\eta=\pi/12$, thus resulting in increasing the maximum of efficiency. We also find that the inclusion of the MOG parameter becomes an increasingly important role in approaching high energy efficiency for the energy extraction via the magnetic reconnection process. From the right panel of Fig.~\ref{fig:efficiency}, the presence of the MOG parameter can influence the energy efficiency. %\cred
{That is, the efficiency of energy extraction via the magnetic reconnection continues to grow and reach its possible maximum at the horizon. This happens because the magnetic reconnection can occur in a slightly larger location $r/M$ under an attractive gravitational effect of the MOG parameter, thus resulting in extracting out the rotational energy {in slightly larger locations in the ergosphere}. It is explicitly shown by Fig.~\ref{fig:efficiency} that the effect of the MOG parameter can shift the location point toward to larger $r/M$. However, it turns out that the higher energy efficiency can be compensated for higher values of the spin parameter for a very nearly extremal black hole case.} Here, the key point to be noted is that the MOG parameter $\alpha$ interprets as an attractive gravitational charge that can physically manifest to strengthen black hole gravity, thus allowing the Kerr-Newman-MOG black hole to have the spin parameter $a>1$ greater than that of the Kerr black hole spin. It does also increase the black hole horizon. The energy efficiency is therefore higher than the one for the Kerr black hole for non-extremal cases, as seen in the right panel of Fig.~\ref{fig:efficiency}. One can then deduce that an attractive gravitational property of the MOG parameter and its combined effect with black hole charge can make the efficiency of energy extraction become significantly efficient through the magnetic reconnection. 

Finally we turn to compare the power of the magnetic reconnection and Blandford-Znajek mechanisms in order to estimate the rate of energy extraction under the fast magnetic reconnection. To that, we first consider the rate of energy extraction for the BZ mechanism at the horizon, and it is defined by  \cite{Tchekhovskoy10ApJ}
\begin{eqnarray}
 P_{\text{BZ}}=\frac{\kappa}{16\pi} \Phi^2_{BH} \Omega_{H}^2
    \bigg[1+\chi_1 \Omega_{H}^2+\chi_2\Omega_{H}^4+\mathcal{O}(\Omega_{H}^6)\bigg]\, ,
\end{eqnarray}
where $\Phi_{BH}$ and $\Omega_H$, respectively, denote the magnetic flux and the angular velocity $\Omega_{H}$  at the horizon which is given by
\begin{eqnarray}
\Omega_{H}=\frac{a}{2\mathcal{M}r-Q^2-\frac{\alpha}{1+\alpha}\mathcal{M}^2}\, ,
\end{eqnarray}
while $\kappa$, $\chi_1$, and $\chi_2$ refer to numerical constants.  Here we note that $\kappa$ has a relation with geometric configuration \cite{Pei16}. The magnetic flux is then defined by $\Phi_{BH}=\frac{1}{2}\int_{\theta}\int_{\phi}|B^r|dA_{\theta\phi}$, and we shall for simplicity assume $\Phi_{BH}\sim 2\pi B_0\,r_{H}^{2}\,\sin\xi$ for further analysis. Here we do, however, note that for the magnetic field configuration the orientation angle $\xi$ can be regarded as a good estimate only at low latitudes. In fact, one needs to evaluate the magnetic flux $\Phi_{BH}$ with the magnetic field configuration at all latitudes~\cite{Koide02Sci,Semenov04Sci}. Taking all together the ratio of these two powers $P_{MR}/P_{BZ}$ can then be written by \cite{Comisso21}
\begin{eqnarray}
    \dfrac{P_{MR}}{P_{BZ}}=\dfrac{-4\,\epsilon_{-}^{\infty}{\mathit{w}}_0 A_{in} U_{in}}{ \pi\kappa\, \Omega^2_H r^4_H \sigma_0 \sin^2{\xi}(1+\chi_1 \Omega^2_H +\chi_2 \Omega^4_H)}\, ,
\end{eqnarray}
where we note that the values of these numerical constants have been approximated as $\chi_1\approx 1.38$ and $\chi_2\approx -9.2$, while $\kappa\approx 0.044$ due to the magnetic field geometry (see, for example, \cite{Pei16}).

In Fig.~\ref{fig:ratio} we show the behavior of $P_{MR}/P_{BZ}$ against the plasma magnetization for various combinations of the location $r/M$ while keeping black hole parameters fixed. From the left panel of Fig.~\ref{fig:ratio}, the height of the ratio of powers decreases and the curves shift right to larger $\sigma_0$ as the location $r/M$ increases. It does, however, satisfy $P_{MR}/P_{BZ}>1$ that makes the magnetic reconnection significantly more efficient than the one for the BZ mechanism.  Similar to what is observed in the left panel of Fig.~\ref{fig:ratio}, the right panel reflects the impact of the combined effect of black hole charge and MOG parameter on the power ratio $P_{MR}/P_{BZ}$ for different possible locations $r/M$. As the MOG parameter increases, the curves of the power ratio shift left to smaller $\sigma_0$ for both locations, as seen in the right panel of Fig.~\ref{fig:ratio}. It can be seen that the shape of the ratio $P_{MR}/P_{BZ}$ increases due to the combined effect of black hole charge and MOG parameter, thus resulting in increasing the efficiency of magnetic reconnection and allowing to mine out more energy from the black hole than for the BZ mechanism.

\section{Conclusions}
\label{Sec:conclusion}

Black holes are huge reservoirs of energy and, for this reason, the enormous luminosity of AGN is connected to the supermassive black holes anchored in the centers of galaxies. It is believed that the rotational energy is the most important aspect of astrophysical rotating black holes as their own reducible energy, as predicted by GR. It is therefore fundamentally important to understand more deeply this energy sourcing genesis of the highly powerful astrophysical phenomena occurring in the vicinity of a black hole. 

To that, proposed explanations for these highly powerful astrophysical phenomena come into play in order to mine out black hole rotational energy by assuming the magnetic field existing on the surrounding environment of the black hole. One of them is the Blandford and Znajek mechanism \cite{Blandford1977} that was used to extract the energy using an electromagnetic field existing around the black hole. It is envisaged by the fact that the BZ mechanism utilizes twisting of magnetic field lines because of the frame dragging effect around a rotating black hole which creates a potential difference $U$ and electric current $I$ that comes from the discharging. As a consequence of this process, rotational energy $W\sim IU$ could be mined out from a spinning black hole. 

Another interesting mechanism is the magnetic Penrose process proposed to drive black hole rotational energy out \cite{Bhat85,Parthasarathy86}. This mechanism is also an extremely effective mechanism for extracting energy from rotating black holes through the influence of the magnetic field. This happens because the magnetic field plays an important key role to generate the Wald electric charge of the Kerr black hole which is acting as an accelerator for charged particles in the ergosphere.

The magnetic reconnection mechanism has been recently proposed as the most promising mechanism for extracting out the rotational energy of astrophysical black holes. This mechanism is surprisingly highly efficient,  originally, for extracting out the rotational energy of standard the Kerr black hole by the magnetic reconnection known as the source of magnetar's enormous energetics. It is driven by the fact that the magnetic reconnection burns magnetic field energy for the plasma acceleration through drastic reconfiguration of magnetic field lines due to the frame dragging effect of a rapidly spinning black hole. 

In this paper, we explored the impact of the rotating Kerr-Newman-MOG black hole on the magnetic reconnection by adapting the recently developed Comisso-Asenjo mechanism \cite{Comisso21} and studied the energy extraction by the magnetic reconnection process that occurs continuously inside the ergosphere due to black hole spin. We further analyzed the energy efficiency of energy extraction and the power as a function of plasma magnetization, magnetic field orientation and black hole spin and MOG parameters by imposing all required conditions.  

We explored the accelerated $\epsilon_{+}^{\infty}>0$ and decelerated $\epsilon_{-}^{\infty}<0$ energies of the plasma for the energy extraction through the magnetic reconnection and demonstrated that the accelerated and decelerated energies per enthalpy grow as a consequence of combined effect of the MOG parameter and black hole charge, thus leading to arbitrarily high energy that can be extracted out by the magnetic reconnection. We also studied the phase-space region ($a, r/M$) for satisfying  the required condition for energy extraction. We found that the combined effect of black hole charge and MOG parameter can extend the phase space for the energy extraction condition which leads to arbitrarily high energy driven out from Kerr-Newman-MOG black hole via magnetic reconnection. 

Further, we studied the power and the energy efficiency of a rapidly spinning Kerr-Newman-MOG black hole in order to understand how the magnetic reconnection is efficient as compared to the Kerr black hole case. We showed that  the combined effect of black hole charge $Q$ and MOG parameter $\alpha$ leads to high power through the magnetic reconnection as compared to the Kerr black hole. We also found that the combined effect of MOG parameter and black hole charge can play an increasingly important role in approaching high energy efficiency for energy extraction, i.e., the efficiency of energy extraction reaches up to its possible maximum due to the fact that the MOG parameter $\alpha$ as an attractive gravitational charge can physically manifest to strengthen black hole gravity which can lead to the spin parameter $a>1$ and to high energy efficiency via the magnetic reconnection.

Also, we estimated the rate of energy extraction under the fast magnetic reconnection by comparing the power of the magnetic reconnection and Blandford-Znajek mechanisms. For that, we analyzed the behavior of $P_{MR}/P_{BZ}$ against the plasma magnetization for various possible cases. We showed that the ratio $P_{MR}/P_{BZ}$ increases as a consequence of the combined effect of black hole charge and MOG parameter. Hence, it suggests that the magnetic reconnection is significantly more efficient than for the BZ mechanism and it results in increasing the efficiency as compared to the Kerr black hole. 

From the present results, one can infer that the Kerr-Newman-MOG black hole's impact on the magnetic reconnection is significantly more efficient to extract the rotational energy from the black hole. In fact, magnetic reconnection is fueled by magnetic field energy due to the twisting of magnetic field lines around the black hole for plasma acceleration. In this regard, the MOG parameter can cause even more fast spin that strongly affects the reconfiguration of magnetic field lines due to the frame dragging effect. This is how the combined effect of black hole charge and MOG parameter makes the energy extraction surprisingly more efficient for the Kerr-Newman-MOG black hole as compared to the Kerr black hole.

\section*{Acknowledgments}
We warmly thank Luca Comisso for valuable comments and discussions that definitely helped to improve the accuracy and quality of the presentation of the manuscript. We also thank Pankaj Sheoran for useful discussions. This work is supported by the National Natural Science Foundation of China under Grant No. 11675143 and the National Key Research and Development Program of China under Grant No. 2020YFC2201503. M.A. and B.A. wish to acknowledge the support from Research Grant No. F-FA-2021-432 of the Uzbekistan Agency for Innovative Development.

\appendix

\bibliographystyle{apsrev4-1}  %% BibTeX style
\bibliography{gravreferences,ref}

\end{document}